\documentclass[useAMS,usenatbib]{mn2e}
\usepackage{amsmath}
\usepackage{epsfig}
\usepackage{txfonts}

\usepackage[pdftex,pdfpagemode={UseOutlines},bookmarks,bookmarksopen,colorlinks,linkcolor={blue},citecolor={blue},urlcolor={red}]{hyperref}
\usepackage{float}
\usepackage{graphicx}
\usepackage{caption}
\usepackage{subcaption}
\usepackage{natbib}
\usepackage{rotating,longtable,lscape}

\newcommand{\twopartdef}[4]
{
	\left\{
		\begin{array}{ll}
			#1 & \mbox{if } #2 \\ \\
			#3 & \mbox{if } #4
		\end{array}
	\right.
}

\title[Analytic circumbinary transit probability]{Circumbinary planets - why they are so likely to transit}

\author[Martin and Triaud]
{\parbox{\textwidth}{David. V. Martin$^1$\thanks{E-mail: david.martin@unige.ch}, Amaury. H.M.J. Triaud$^{2,3,4}$}
\vspace{0.4cm}\\
\parbox{\textwidth}{$^{1}$Observatoire de Gen\`eve, Universit\'e de Gen\`eve, 51 chemin des Maillettes, Sauverny 1290, Switzerland\\
$^{2}$Department of Physics, and Kavli Institute for Astrophysics and Space Research, Massachusetts Institute of Technology, Cambridge, MA 02139, USA\\
$^{3}$Centre for Planetary Sciences, University of Toronto, Department of Environmental and Physical Sciences, 1265 Military Trail, Toronto, Ontario M1C 1A4, Canada\\\
$^{4}$Fellow of the Swiss National Science Foundation}}

\begin{document}

\date{Accepted . Received}

\pagerange{\pageref{firstpage}--\pageref{lastpage}} \pubyear{2014}

\maketitle

\label{firstpage}

\begin{abstract}

Transits on single stars are rare. The probability rarely exceeds a few per cent. Furthermore, this probability rapidly approaches zero at increasing orbital period. Therefore transit surveys have been predominantly limited to the inner parts of exoplanetary systems. Here we demonstrate how circumbinary planets allow us to beat these unfavourable odds. By incorporating the geometry and the three-body dynamics of circumbinary systems, we analytically derive the probability of transitability, a configuration where the binary and planet orbits overlap on the sky. We later show that this is equivalent to the transit probability, but at an unspecified point in time. This probability, at its minimum, is always higher than for single star cases. In addition, it is an increasing function with mutual inclination. By applying our analytical development to eclipsing binaries, we deduce that transits are highly probable, and in some case guaranteed. For example, a circumbinary planet revolving at 1 AU around a 0.3 AU eclipsing binary is certain to eventually transit - a 100\% probability - if its mutual inclination is greater than $0.6^{\circ}$. We show that the transit probability is generally only a weak function of the planet's orbital period;  circumbinary planets may be used as practical tools for probing  the outer regions of exoplanetary systems  to search for and detect warm to cold transiting planets. 

\end{abstract}


\section{Introduction}
\label{sec:intro}

In the burgeoning search for extra-solar planets, circumbinary planets represent some of the most exotic systems found to date. They pose astronomers with interesting questions regarding their detectability \citep{schneider94}, abundance \citep{armstrong14,martin14}, formation \citep{pierens13,kley14}, habitability \citep{haghighipour13,mason14}, orbital dynamics \citep{leung13} and stability \citep{dvorak86,dvorak89,holman99}.

Answers to these questions are reliant on planet detections. So far there have been reported discoveries from several techniques, including PSR B1620-26 (\citealt{thorsett99} with pulsar timing), HD 202206 (\citealt{correia05} with radial velocimetry), DP Leonis (\citealt{qian10} with eclipse timing variations), Ross 458 (\citealt{burgasser10} with direct imaging), and Kepler-16 (\citealt{doyle11} with transit photometry). There are presently ten transiting circumbinary planets known, all found by the {\it Kepler} telescope \citep{welsh14}.

The advantage of finding circumbinary planets in transit is that they can yield an unambiguous detection, thanks to a unique signature that is hard to mimic with false positives. The photometric measurement of the radius can be complemented with transit timing variations (TTVs), eclipse timing variations (ETVs) or spectroscopy to obtain the mass and bulk density, which are important from a formation perspective. Transits also open the door to atmospheric characterisation \citep{seager10}, the measure of the Rossiter-McLaughlin effect \citep{queloz00,fabrycky14}, and the detection of exomoons \citep{kipping12}.

It will be shown in this paper that circumbinary planets, beyond their exoticity, are useful astronomical tools. Their particular geometry and orbital dynamics lead to potentially much higher transit probabilities in comparison with single stars. There is also a weaker dependence on orbital period, allowing us to extend transit studies to the outer regions of stellar systems. 

The paper is structured as follows. In Sect.~\ref{sec:geometry} we introduce the geometry of circumbinary planets. Next in Sect.~\ref{sec:dynamics} we analyse the orbital dynamics of circumbinary systems and the effects on their observability. We then define the concept of transitability and analytically derive a criterion for its occurrence in Sect.~\ref{sec:criterion}. Following this, we convert this criterion into the probability of a circumbinary system exhibiting transitability in Sect.~\ref{sec:probability}, similar to the work done for single stars \citep{borucki84,barnes07}. In Sect.~\ref{sec:EBs} we analyse the special case of eclipsing binaries.

As an observer, the observable quantity is a transit, not transitability. This is why in Sect.~\ref{sec:connection} we connect the two concepts, verifying that a system exhibiting transitability is effectively guaranteed to transit, albeit at an unspecified point in time. Some illustrative transit wait times are calculated, revealing that they may be within a few years for many systems. In Sect.~\ref{sec:discussion} we discuss some applications and limitations of our work, before concluding in Sect.~\ref{sec:conclusion}.




\section{geometry}
\label{sec:geometry}

\begin{figure}  
\begin{center}  
\includegraphics[width=0.5\textwidth]{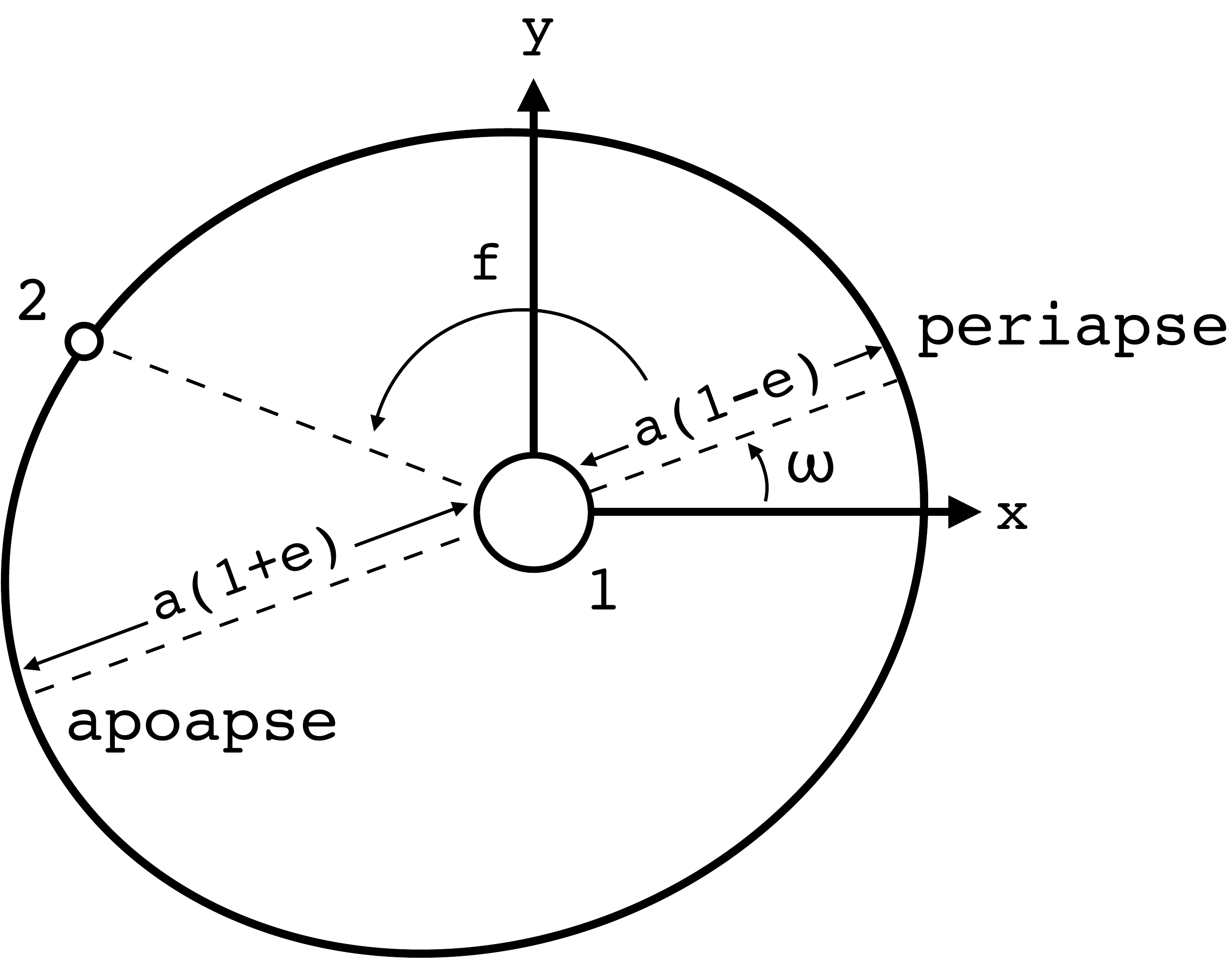}  
\caption{Planar orbital elements of a two-body system.}
\label{fig:2DOrbit}
\end{center}  
\end{figure}

\begin{figure}  
\begin{center}  
\includegraphics[width=0.5\textwidth]{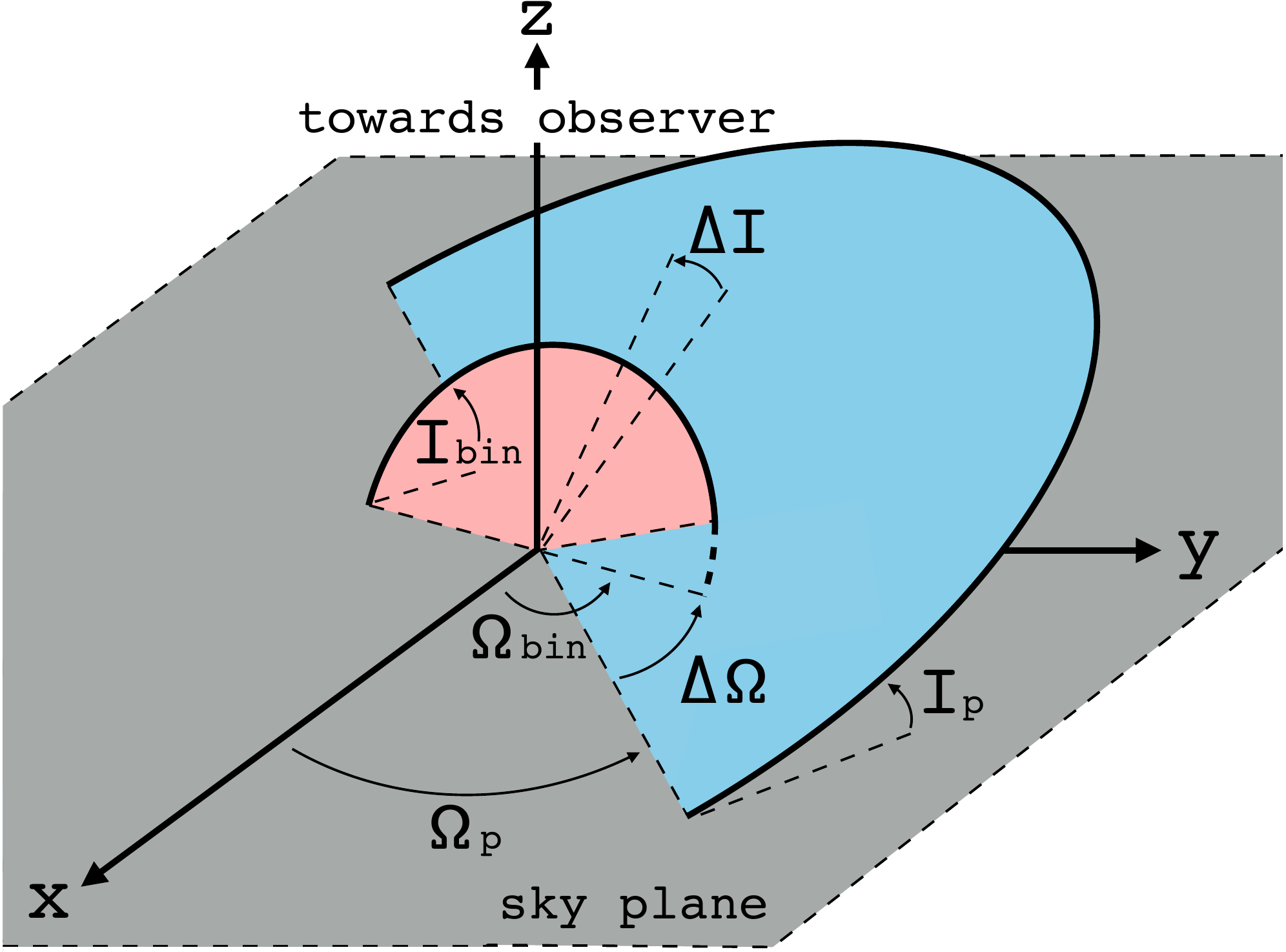}  
\caption{A circumbinary planet in a misaligned orbit (blue, outer) around a binary star system (pink, inner). The misalignment is characterised by the mutual inclination, $\Delta I$, and the mutual longitude of the ascending node, $\Delta \Omega$. The observer is looking down the z-axis from above, and hence the grey x-y plane denotes the plane of the sky.}
\label{fig:MisalignedCircumbinarySystem}
\end{center}  
\end{figure}

We will treat a circumbinary system as a pair of Keplerian orbits in Jacobi coordinates, with the addition of first-order dynamical effects (Sect.~\ref{sec:dynamics}). The inner orbit is the stellar binary (subscript ``bin"). The outer orbit is the planet around the binary centre of mass (subscript ``p"). Each Keplerian orbit is an ellipse characterised by four orbital elements: the semi-major axis $a$, eccentricity $e$, argument of periapsis $\omega$ and true anomaly $f$. These quantities are defined in Fig.~\ref{fig:2DOrbit}. This set of four is not unique, and often we will use the period $T$ instead of the semi-major axis. The orientation of each orbit in three dimensions is defined using two extra angles: the inclination $I$ and longitude of the ascending node $\Omega$. In Fig.~\ref{fig:MisalignedCircumbinarySystem} we depict the 3D orientation of a binary and planet orbit. We take the observer to be looking down the z-axis. An eclipsing binary, for example, corresponds to $I_{\rm bin} \approx \pi/2$. Throughout this paper we use radians unless otherwise specified with a $^{\circ}$ symbol.

The orientation of the planetary orbit with respect to the binary is characterised by two quantities: the mutual inclination

\begin{equation}
\label{eq:MutualInclCalcComplete}
\cos\Delta I = \sin I_{\rm bin}\sin I_{\rm p}\cos\Delta \Omega + \cos I_{\rm bin}\cos I_{\rm p},
\end{equation}
and the mutual longitude of the ascending node

\begin{equation}
\label{eq:DeltaOmega}
\Delta \Omega = \Omega_{\rm bin} - \Omega_{\rm p},
\end{equation}
which are also shown in Fig.~\ref{fig:MisalignedCircumbinarySystem}. When using transit photometry or radial velocimetry, the observer is sensitive to $\Delta \Omega$ but not to the individual quantities $\Omega_{\rm bin}$ and $\Omega_{\rm p}$. Throughout this paper can we therefore take $\Omega_{\rm bin}=0$ and allow $\Omega_{\rm p}$ to vary.

\section{Dynamic orbits}
\label{sec:dynamics}

\begin{figure}  
\begin{center}  
\includegraphics[width=0.5\textwidth]{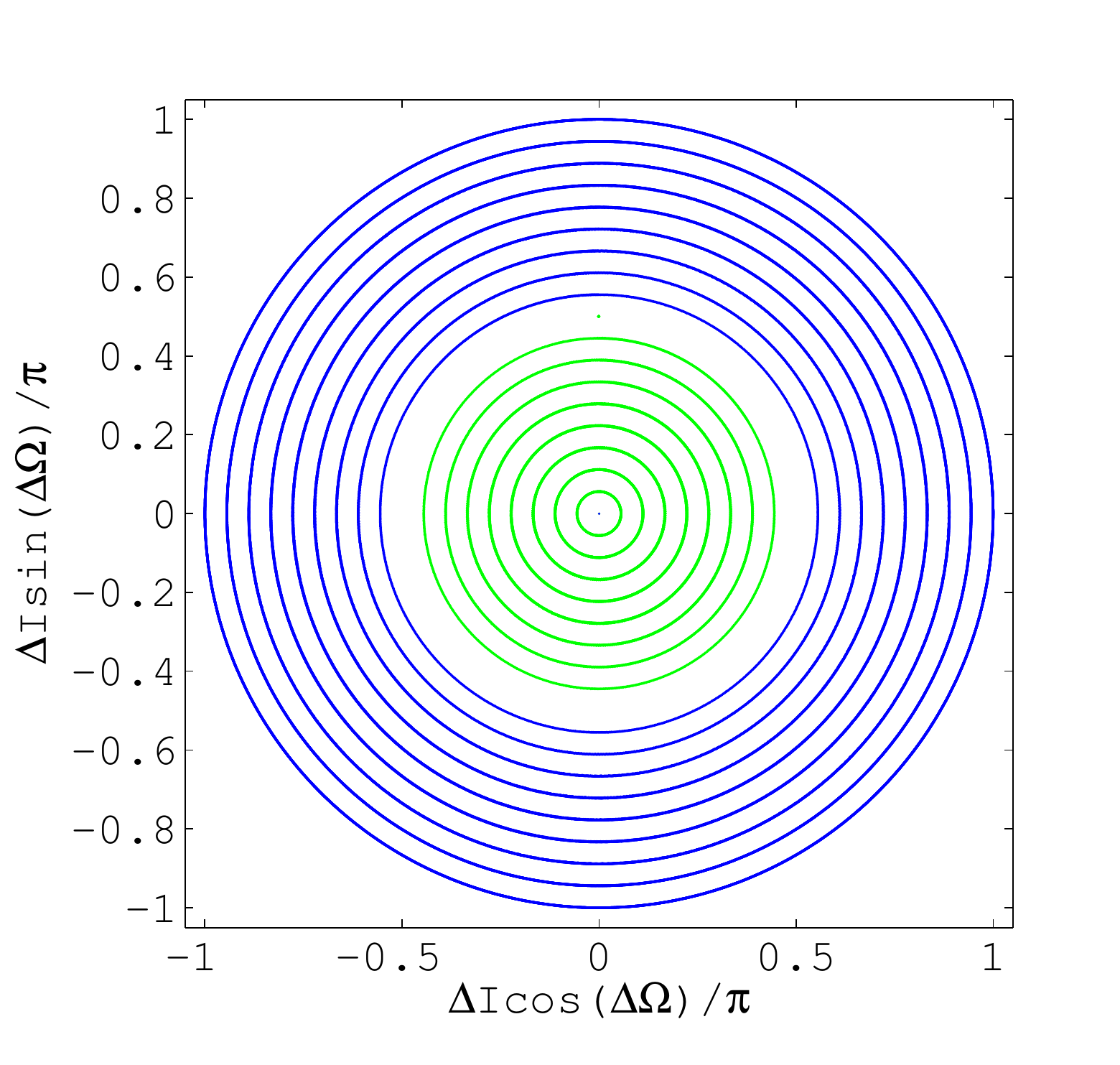}  
\caption{Surfaces of section of the mutual inclination, $\Delta I$, and mutual longitude of the ascending node, $\Delta  \Omega$, between the binary and planet orbital planes.}
\label{fig:DoolinBlundellTest}
\end{center}  
\end{figure} 

A static Keplerian orbit is insufficient for accurately describing a circumbinary planet. Owing to perturbations from the binary, the orbital elements defined in Sect.~\ref{sec:geometry} vary on observationally relevant timescales. We include in our derivation the most prominent of these effects: a precession in the planet's orbital plane. This behaviour is described by a time-variation in $\Delta I$, and $\Delta \Omega$. We restrict ourselves to circular binaries and planets\footnote{It is technically a misnomer to speak of circular circumbinary orbits, since perturbations from the binary cause $e_{\rm p}$ to vary even if initially zero. However, this only has a very small effect on the transit geometry. See Sect.~\ref{sec:additional_effects} for further detail.}. In this case, the orbital plane of the planet rotates at a constant rate around the normal to the binary plane ($\Delta \Omega  = 0$ to $2\pi$), whilst maintaining $\Delta I = {\rm const}$ \citep{schneider94,farago10,doolin11}. The precession period $T_{\rm prec}$ according to \citet{schneider94} is

\begin{equation}
\label{eq:schneider_prec_period}
T_{\rm prec} = T_{\rm p}\frac{16}{3}\left(\frac{a_{\rm p}}{a_{\rm bin}}\right)^2\frac{1}{\cos \Delta I},
\end{equation}
where the stars are assumed to be of equal mass. An alternative, more complex derivation can be found in \citet{farago10}. The planet orbit is stable as long as it is not too close to the binary. An approximate criterion from the work of \citet{dvorak86,dvorak89,holman99} is

\begin{equation}
\label{eq:stability_criterion}
a_{\rm p} \gtrsim 3 a_{\rm bin}.
\end{equation}

In Fig.~\ref{fig:DoolinBlundellTest} we demonstrate how precession effects $\Delta I$ and $\Delta \Omega$ using numerical N-body integrations\footnote{All N-body simulations in this paper are done using a fourth-order Runge-Kutta algorithm, where energy loss due to its non-symplectic nature was kept to negligible levels.}. We ran a set of simulations with $\Delta \Omega$ starting at $90^{\circ}$ and $\Delta I$ varied between 0$^{\circ}$ and 180$^{\circ}$ in steps of 10$^{\circ}$ and, each corresponding to a different curve in Fig.~\ref{fig:DoolinBlundellTest}. The stars are of mass $1 M_{\odot}$ and $0.5 M_{\odot}$ with $a_{\rm bin}=0.07$ AU. The planet is a massless test particle with $a_{\rm p} = 0.3$ AU. The green, inner curves are for prograde orbits with clockwise precession. The blue, outer curves are for retrograde orbits with anti-clockwise precession. The gap between the green and blue curves corresponds to $\Delta I = 90^{\circ}$, i.e. for a polar orbit where the precession period becomes infinitely long (Eq.~\ref{eq:schneider_prec_period}).

\begin{figure}  
\begin{center}  
\includegraphics[width=0.5\textwidth]{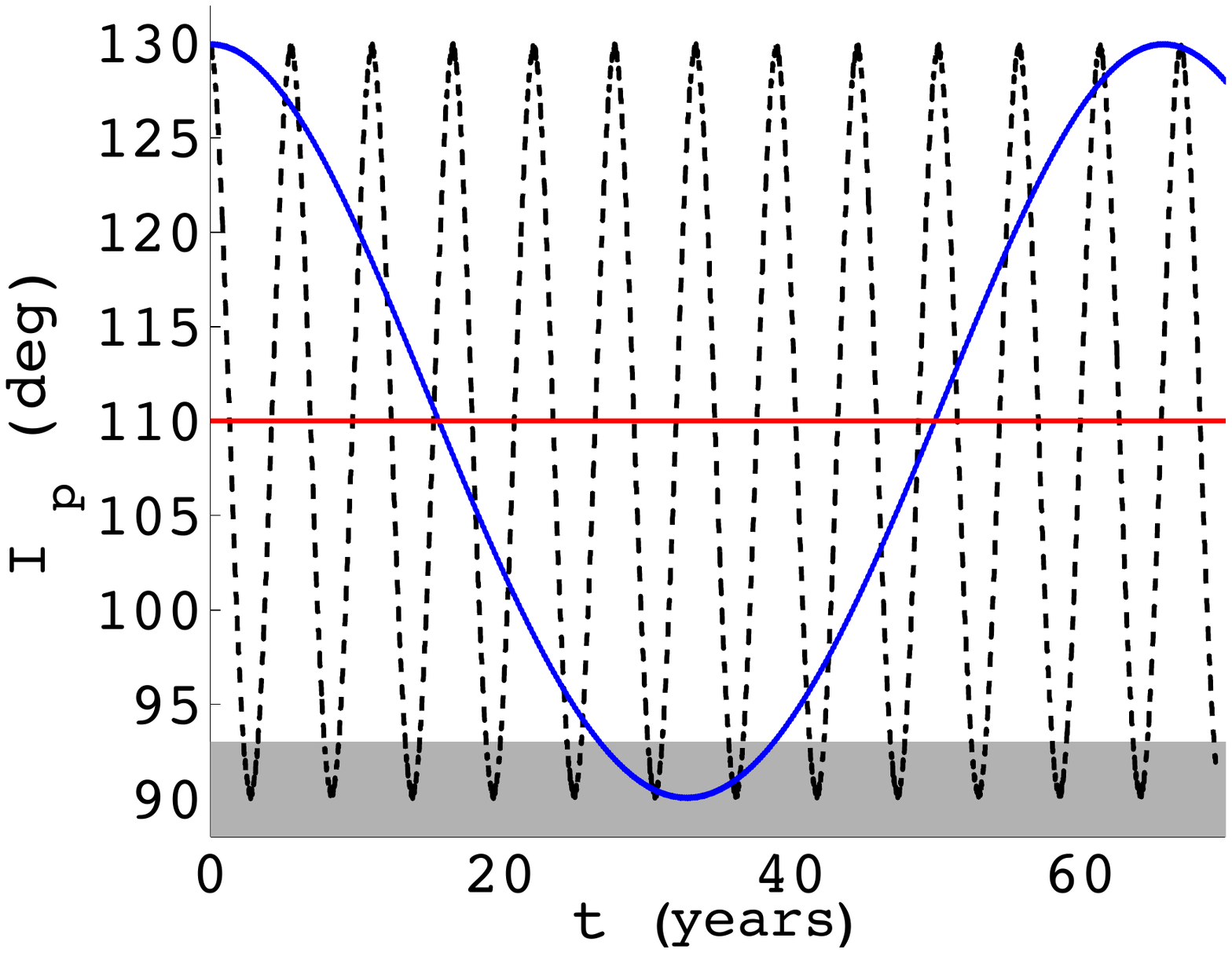}  
\caption{Variation of $I_{\rm p}$ over time for two circumbinary systems ($a_{\rm p}=0.3$ AU in black and white dashes, $a_{\rm p} = 0.6$ AU in blue). The horizontal red line denotes the binary's orbital plane inclination on the sky, $I_{\rm bin}$. The grey region corresponds to when the planet is in transitability.}

\label{fig:SkyInclinationTest}
\end{center}  
\end{figure}

These orbital dynamics have observational consequences. The inclination planet on the sky, $I_{\rm p}$ varies with time according to
\begin{equation}
\label{eq:PlanetSkyInclination}
I_{\rm p} = \Delta I \cos\left(\frac{2\pi}{T_{\rm prec}}t\right) + I_{\rm bin},
\end{equation}
where $t$ is time. An example is shown in Fig.~\ref{fig:SkyInclinationTest} for two circumbinary systems. The binary in both systems has equal mass stars with $M_{\rm A} =M_{\rm B}=1 M_{\odot}$, $a_{\rm bin} = 0.1$ AU, $I_{\rm bin} = 110^{\circ}$, and $\Omega_{\rm bin} = 0^{\circ}$. The planet is a massless body with starting values $I_{\rm p}=130^{\circ}$ and $\Omega_{\rm p} = 0^{\circ}$. The mutual inclination is $20^{\circ}$. The two planets shown in the figure have different values for $a_{\rm p}$: $0.3$ AU for the black, dashed sinusoid and $0.6$ AU for the blue, solid sinusoid.

The maximum and minimum values of $I_{\rm p}$ are independent of $a_{\rm p}$. The planet semi-major axis does, however, strongly influence the precession period: $T_{\rm prec} = 5.6$ yr for $a_{\rm p} = 0.3$ AU and 65.9 yr for $a_{\rm p} = 0.6$ AU, according to the N-body simulation. The analytic expressions in Eq.~\ref{eq:schneider_prec_period} produces precession periods of 6.0 yr and 67.1 yr, showing it to be reasonably accurate.

%

\section{Criterion for transitability}
\label{sec:criterion}


\begin{figure}  
\begin{center}  
\includegraphics[width=0.5\textwidth]{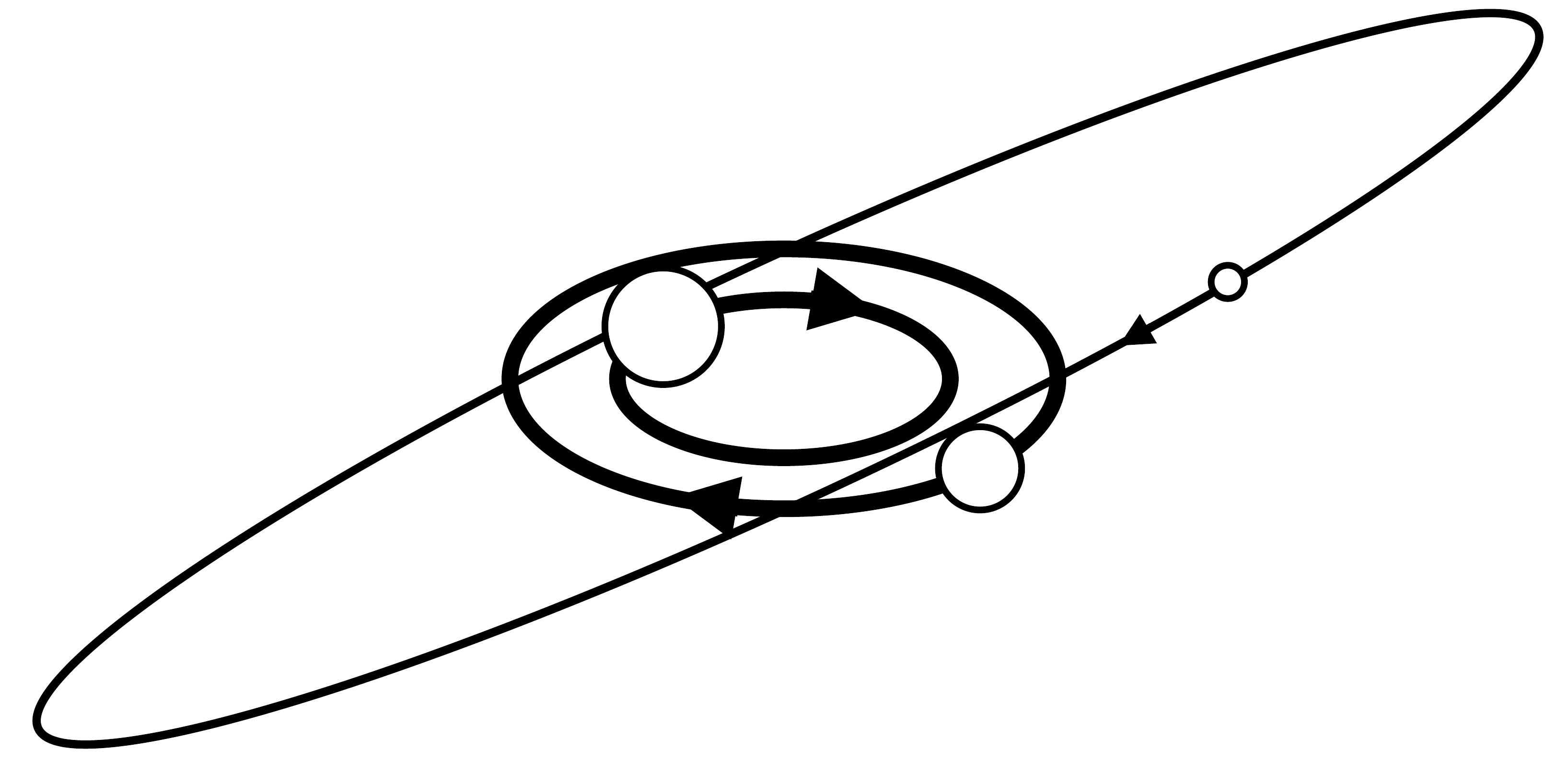}  
\caption{An example circumbinary system exhibiting transitability.}
\label{fig:TransitabilityExample}
\end{center}  
\end{figure} 

Transitability is an orbital configuration where the planet and binary orbits intersect on the sky, like the example shown in Fig.~\ref{fig:TransitabilityExample}. In this scenario transits are possible but not guaranteed on every passage of the planet past the binary, because of the relative motion of the three bodies. This terminology was first introduced in \citet{martin14}, we where formally defined and elaborated upon a concept that had been already used in several studies \citep{schneider94,welsh12,kratter13}. In Fig.~\ref{fig:SkyInclinationTest} the grey region denotes the time spent in transitability. During each precession period, there will be zero, one or two intervals of transitability, or permanent transitability in only two scenarios: 1) $I_{\rm bin}$ and $I_{\rm p}$ are both very close to $\pi/2$ and 2) polar orbits where $I_{\rm bin}=0$ and $I_{\rm p} =\pi/2$.

We work to derive a criterion that predicts whether or not a circumbinary planet will enter transitability at any point during the precession period. As a first approximation, we know that the planet is in transitability when the planet orbit is perpendicular to the plane of the sky:

\begin{equation}
\label{eq:transitability_definition_firstorder}
I_{\rm p}=\frac{\pi}{2}.
\end{equation}
This is the most conservative case possible, since it ignores the finite extent of the binary. According to Eq.~\ref{eq:PlanetSkyInclination}, $I_{\rm p}$ is guaranteed to reach $\pi/2$ if


\begin{equation}
\label{eq:TransitabilityInequalityFirstOrder}
\Delta I > \left|\frac{\pi}{2} - I_{\rm bin}\right|.
\end{equation}
This is the first-order criterion for transitability. Three things are apparent: 1) mutual inclinations aid transitability, which is contrary to conventional views on transit geometries, 2) this criterion is independent of the planet period and 3) this criterion is easiest to fulfil at $I_{\rm bin} \approx \pi/2$, i.e. for eclipsing binaries. This criterion was also derived by \citet{schneider94}, who was the first author to analyse circumbinary transit probabilities in the presence of precession.

The second level of complexity is to include the full extent of the stellar orbits, meaning that a value of $I_{\rm p}$ offset from $\pi/2$ may still exhibit transitability. Consider the limiting case of transitability. This is when the planet and binary orbits barely overlap when $|I_{\rm p}-\pi/2|$ is at a minimum ($dI_{\rm p}/dt =0$). We calculate the orientation of the binary and planet orbits in this configuration. Take Eq.~\ref{eq:MutualInclCalcComplete} and rearrange it to isolate the term containing $\Delta \Omega$:

\begin{equation}
\label{eq:Limit_Orientation_1}
\cos \Delta \Omega = \frac{\cos \Delta I - \cos I_{\rm bin}\cos I_{\rm p}}{\sin I_{\rm bin}\sin I_{\rm p}}.
\end{equation}
In Eq.~\ref{eq:Limit_Orientation_1} only $I_{\rm p}$ and $\Delta \Omega$ are time-dependent quantities. Differentiating both sides of Eq.~\ref{eq:Limit_Orientation_1} with respect to time leads to


\begin{align}
\label{eq:Limit_Orientation_3}
\begin{split}
   &-\sin \Delta \Omega \frac{d\Delta \Omega}{dt} = \left[\cos I_{\rm bin}\sin I_{\rm p}\frac{dI_{\rm p}}{dt}\sin I_{\rm bin}\sin I_{\rm p} \right. \\ 
    &-\left. (\cos \Delta I - \cos I_{\rm bin}\cos I_{\rm p})\sin I_{\rm bin}\cos I_{\rm p}\frac{dI_{\rm p}}{dt} \right] \\
   & / (\sin^2 I_{\rm bin} \sin^2 I_{\rm p}).
\end{split}
\end{align}
By substituting $dI_{\rm p}/dt=0$ into Eq.~\ref{eq:Limit_Orientation_3} we get

\begin{equation}
\label{eq:Limit_Orientation_4}
   \sin \Delta \Omega \frac{d\Delta \Omega}{dt} = 0.
\end{equation}
From Sect.~\ref{sec:dynamics} it is known that $d \Delta \Omega / dt = {\rm const} \neq 0$, and hence $\sin \Delta \Omega=0$, implying that $\Delta \Omega=0$. This means that the limiting case of transitability occurs when the ascending nodes of the binary and planet orbits are aligned. This simplifies the geometry and calculations. According to Sect.~\ref{sec:dynamics}, $\Delta \Omega$ is guaranteed to equal zero at some point during the precession period. When $\Delta \Omega = 0$ the mutual inclination calculation in Eq.~\ref{eq:MutualInclCalcComplete} is simplified to

\begin{equation}
\label{eq:MutualInclCalcSimple}
\Delta I = |I_{\rm bin} - I_{\rm p}|.
\end{equation}

\begin{figure}  
\begin{center}  
\includegraphics[width=0.5\textwidth]{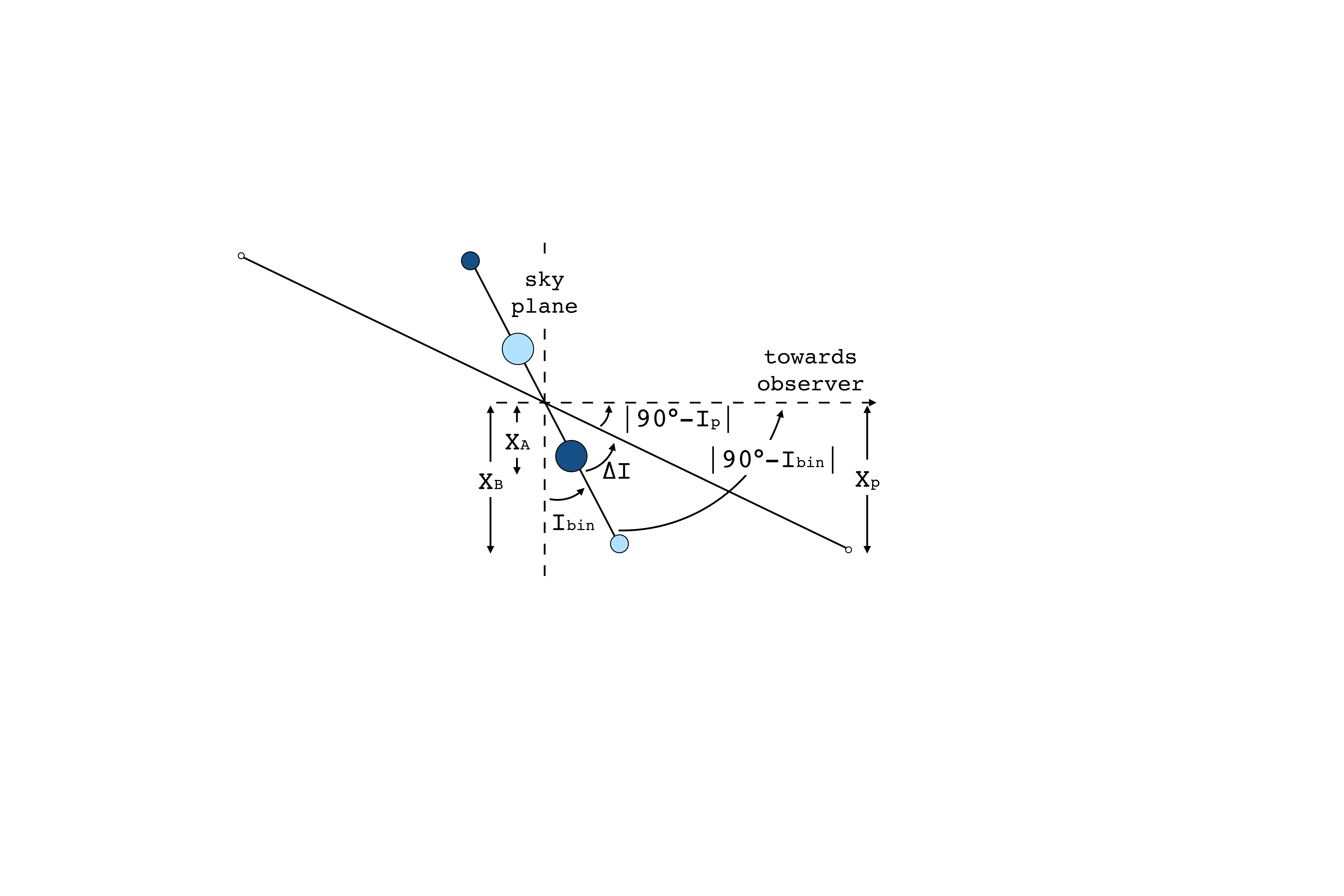}  
\caption{A side-on view of a circumbinary system in the limiting case of transitability, where $\Delta \Omega=0$. The two extreme vertical positions of each star on the sky are drawn in different colours. In this example, the planet is barely in transitability on the secondary star, but not on the primary.}
\label{fig:side_view}
\end{center}  
\end{figure}

In Fig.~\ref{fig:side_view} we show a circumbinary system in the limiting case of transitability. This is a ``side-on" view of the orbit, with the observer located to the right of the page. Using this diagram, we define half the projected heights on the sky of the two stellar orbits to be

\begin{equation}
\label{eq:Xbin}
X_{\rm A,B} = a_{\rm A,B}\sin\left|I_{\rm bin} - \frac{\pi}{2}\right| + R_{\rm A,B},
\end{equation}
where
\begin{equation}
\label{eq:semi-major_axes}
a_{\rm A,B} = a_{\rm bin}\mu_{\rm B,A}
\end{equation}
are the individual semi-major axes for the two stars and
\begin{equation}
\mu_{\rm B,A}=\frac{M_{\rm B,A}}{M_{\rm A}+M_{\rm B}}
\end{equation}
are the reduced masses. Similarly for the planet,
\begin{equation}
\label{eq:Xp}
X_{\rm p} = a_{\rm p}\sin\left|I_{\rm p} - \frac{\pi}{2}\right|.
\end{equation}
There is transitability on stars A and/or B when

\begin{equation}
\label{eq:Xbin=Xp}
X_{\rm p} < X_{\rm A,B},
\end{equation}
According to the orbital dynamics (Eq.~\ref{eq:PlanetSkyInclination}), the planet will enter transitability at some point (fulfilling Eq.~\ref{eq:Xbin=Xp}) if the following criterion is met:

\begin{equation}
\label{eq:beta_factor_inequality}
\Delta I > \left|\frac{\pi}{2} - I_{\rm bin}\right| - \beta_{\rm A,B}(I_{\rm bin}),
\end{equation}
where the angle $\beta_{\rm A,B}$ is a function of the binary extent on the sky, as seen by the planet. This is the second-order criterion for transitability. Combine Eqs.~\ref{eq:MutualInclCalcSimple} and \ref{eq:beta_factor_inequality} to get

\begin{equation}
\label{eq:removed_deltaI}
\left|I_{\rm bin} - I_{\rm p}\right| > \left|\frac{\pi}{2} - I_{\rm bin}\right| - \beta_{\rm A,B}(I_{\rm bin}).
\end{equation}
By inserting Eqs.~\ref{eq:Xbin} and \ref{eq:Xp} into Eq.~\ref{eq:Xbin=Xp} and rearranging to match the form of Eq.~\ref{eq:removed_deltaI}, we obtain



\begin{equation}
\label{eq:beta_A}
\beta_{\rm A,B}(I_{\rm bin}) = \sin^{-1}\left[\frac{a_{\rm A,B}}{a_{\rm p}}\sin\left|\frac{\pi}{2}-I_{\rm bin}\right|+\frac{R_{\rm A,B}}{a_{\rm p}}\right],
\end{equation}
where the quantities inside the square brackets are sufficiently small that we can use the small angle approximation.



The size of $\beta_{\rm A,B}$ determines how easy it is for a system to fulfil the transitability criterion in Eq.~\ref{eq:beta_factor_inequality}. Depending on the separate values of $\beta_{\rm A}$ and $\beta_{\rm B}$, it is possible to fulfil the criterion for just one of the stars. Generally $\beta_{\rm B} > \beta_{\rm A}$ except for eclipsing binaries.

To test the validity of the second-order transitability criterion we ran numerical N-body simulations on thousands of hypothetical circumbinary systems. The details are shown in Appendix~\ref{app:test_criterion}. The analytic criterion is shown to be very accurate, with an error less than 0.1\%. All error cases were near the limit of the inequality in Eq.~\ref{eq:beta_factor_inequality}. Errors arise due to small variations in the semi-major axis and eccentricity, which in the limiting case of transitability may lead to a contrary result to the prediction of Eq.~\ref{eq:beta_factor_inequality}. This is elaborated upon in Sect.~\ref{sec:additional_effects}.

\section{Probability of transitability}
\label{sec:probability}

For a given set of orbital parameters, we can calculate the probability that a given observer will observe transitability, at some time during the precession period. As an initial approximation, we use the first-order transitability criterion (Eq.~\ref{eq:TransitabilityInequalityFirstOrder}). The orientation of a circumbinary system on the sky is uniformly random. It therefore follows that $\cos I_{\rm bin}$ has a uniform distribution, and hence the probability density function is

\begin{equation}
\label{eq:BinaryProbDensityFunction}
p(I_{\rm bin}) = \sin I_{\rm bin}.
\end{equation}
By integrating this between the bounds specified by Eq.~\ref{eq:TransitabilityInequalityFirstOrder}, we obtain an approximate probability of transitability:

\begin{align}
\label{eq:prob_transitability_first_order}
P_{\rm A,B} &= \int_{\pi/2 - \Delta I}^{\pi/2}\sin I_{\rm bin} dI_{\rm bin} \notag \\
&= \sin(\Delta I).
\end{align}
The probability is {\it period-independent} and {\it non-zero}, except for strictly coplanar systems.

\begin{figure}  
\begin{center}  
	\begin{subfigure}[b]{0.49\textwidth}
		\caption{}
		\includegraphics[width=\textwidth]{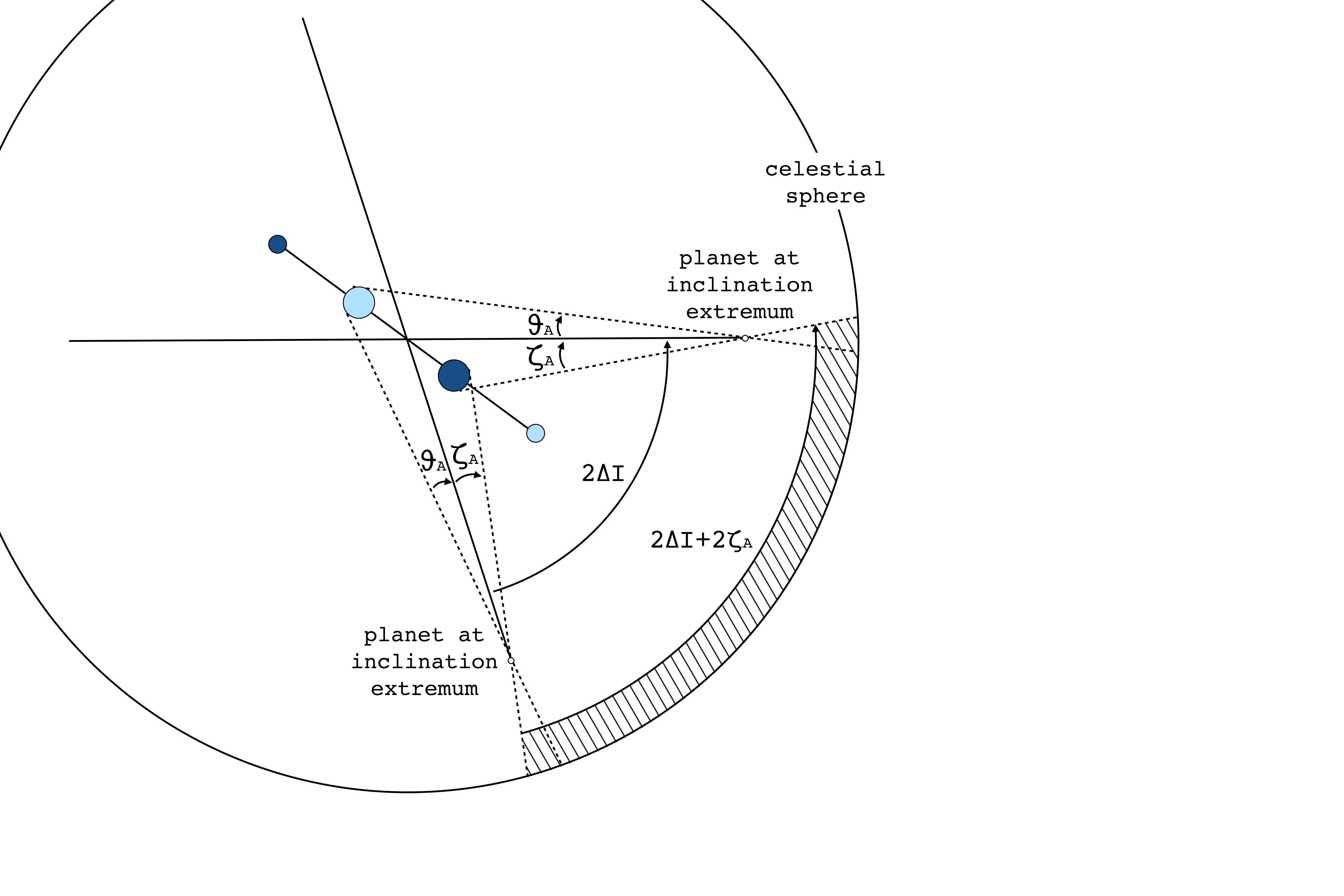}  
		\label{fig:solid_angle_main}  
	\end{subfigure}
	\begin{subfigure}[b]{0.49\textwidth}
		\caption{}
		\includegraphics[width=\textwidth]{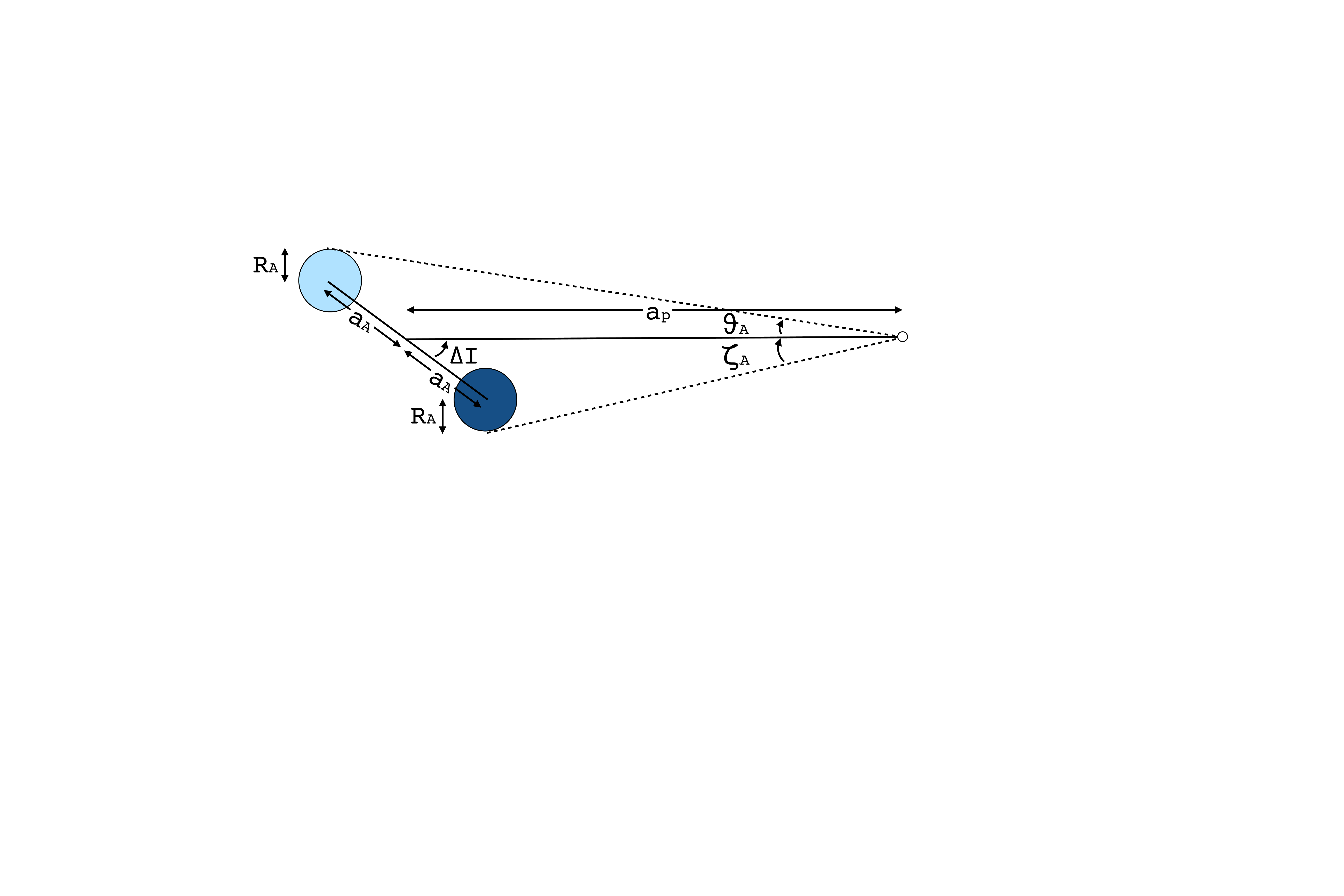}  
		\label{fig:solid_angle_zoom}  
	\end{subfigure}
	\caption{In (a) we show a side-on view of a circumbinary system with the planet shown at its two extreme values of $I_{\rm p}$. Both stars are plotted twice in different colours at closest and farthest separation from the planet. The angles subtended by the orbit on the celestial sphere are only shown for the primary star, to avoid clutter. The hatched region corresponds to observers who would see the planet in transitability at some point in time. In (b) we zoom in to see how $\theta$ and $\zeta$ are defined.
}\label{fig:solid_angle}  
\end{center}  
\end{figure}

The next step is to include the finite extent of the binary orbit. The probability of transitability is a function of the size of the solid angle subtended on the celestial sphere such that the planet and stellar orbits are seen overlapping, including the full orbital evolution. We demonstrate this in Fig.~\ref{fig:solid_angle_main}. To simplify the diagram, we only draw the angles for transitability on the primary star, but the calculation proceeds identically for the secondary. The planet orbit is shown in two different positions, corresponding to the two extrema of $I_{\rm p}$ separated by $2\Delta I$ (Eq.~\ref{eq:PlanetSkyInclination}). The angles $\zeta_{\rm A,B}$ and $\theta_{\rm A,B}$ are the angles subtended by the planet in transitability on each star. They are functions of how big the stellar orbit is, as seen by the planet, at closest ($\zeta_{\rm A,B}$) and farthest ($\theta_{\rm A,B}$) separation. In Fig.~\ref{fig:solid_angle_zoom} we zoom in to see how the angles are defined\footnote{The current equations are corrected from typos that appeared in the published version: in Eq.~\ref{eq:zeta} there used to be a $+$ sign in the denominator and in Eq.\ref{eq:theta} there used to be a $-$ sign in the denominator. We are sorry for the error.}

\begin{equation}
\label{eq:zeta}
\zeta_{\rm A,B} = \tan^{-1}\left(\frac{a_{\rm A,B}\sin \Delta I + R_{\rm A,B}}{a_{\rm p} - a_{\rm A,B}\cos \Delta I}\right)
\end{equation}
and

\begin{equation}
\label{eq:theta}
\theta_{\rm A,B} = \tan^{-1}\left(\frac{a_{\rm A,B}\sin \Delta I + R_{\rm A,B}}{a_{\rm p} + a_{\rm A,B}\cos \Delta I}\right).
\end{equation}
The larger angle $\zeta$ is what corresponds to the limit of transitability and hence $\theta$ does not appear in any further equations.

All observers within the hatched band will eventually see transitability. The probability of an observer being within the hatched band in Fig.~\ref{fig:solid_angle_main} is 
\begin{align}
\label{eq:prob_transitability}
P_{\rm A,B} &= \int_{\pi/2 - \Delta I - \zeta}^{\pi/2}\sin I_{\rm bin} dI_{\rm bin} \notag \\
&= \sin(\Delta I + \zeta_{\rm A,B}) \notag \\
&= \sin\left(\Delta I + \frac{a_{\rm A,B}\sin \Delta I + R_{\rm A,B}}{a_{\rm p} - a_{\rm A,B}\cos \Delta I}\right),
\end{align}
where because $\zeta$ is generally small we can apply the small angle approximation to remove the $\tan^{-1}$ function\footnote{Related to the error mentioned in footnote 3, Eq.~\ref{eq:prob_transitability} has been corrected from the published version which incorrectly contained an $+$ sign in the denominator. This error was purely typographical and all results shown in the paper were calculated using the correct equation.}. Equation~\ref{eq:prob_transitability} is the probability of transitability on the primary and/or secondary stars, for a binary of any orientation. The inclusion of $\zeta$ adds a period-dependency that is absent in Eq.~\ref{eq:prob_transitability_first_order}.

\begin{figure*}  
\begin{center}  
	\begin{subfigure}[b]{0.49\textwidth}
		\caption{}
		\includegraphics[width=\textwidth]{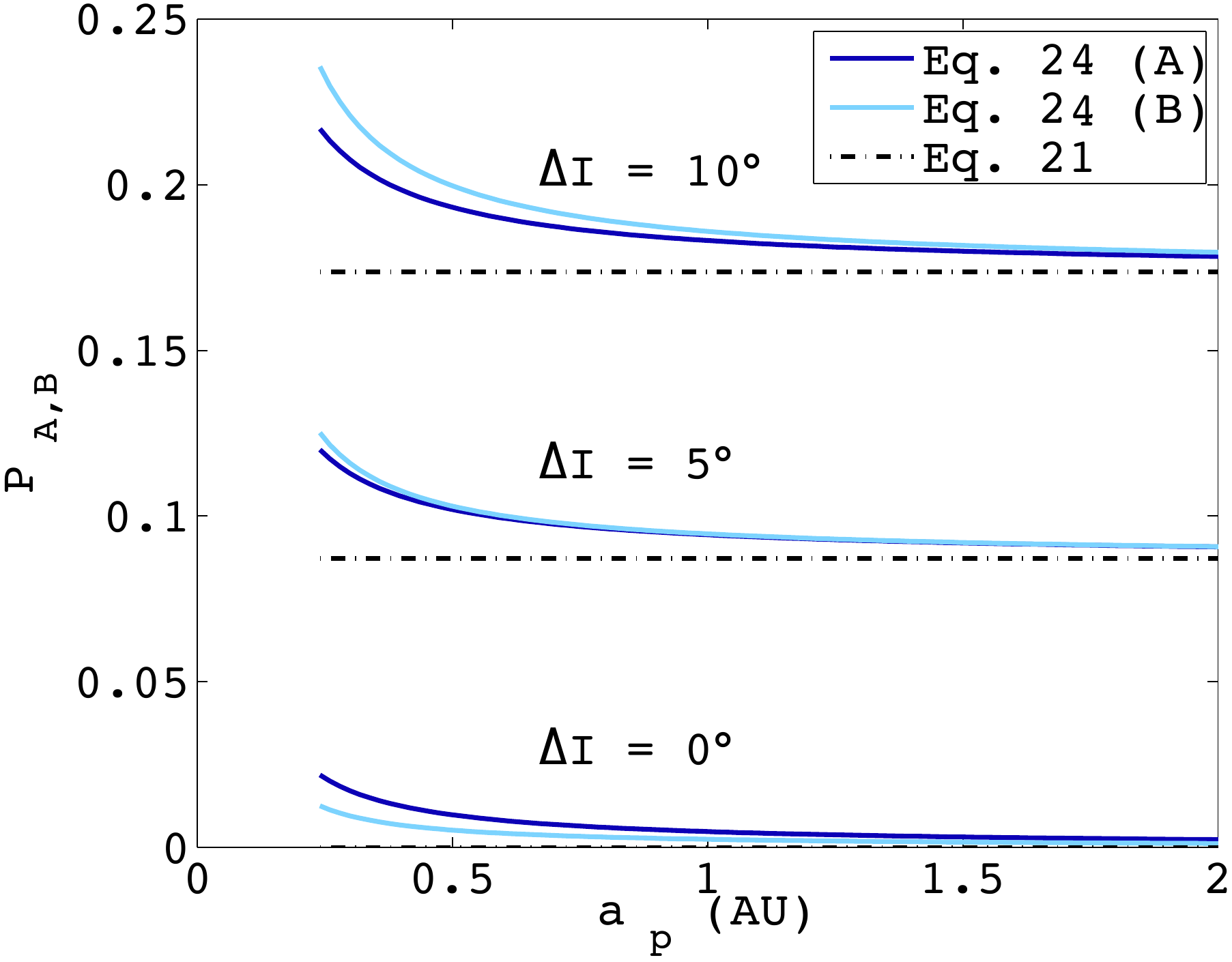}  
		\label{fig:Prob_vs_ap_1}  
	\end{subfigure}
	\begin{subfigure}[b]{0.49\textwidth}
		\caption{}
		\includegraphics[width=\textwidth]{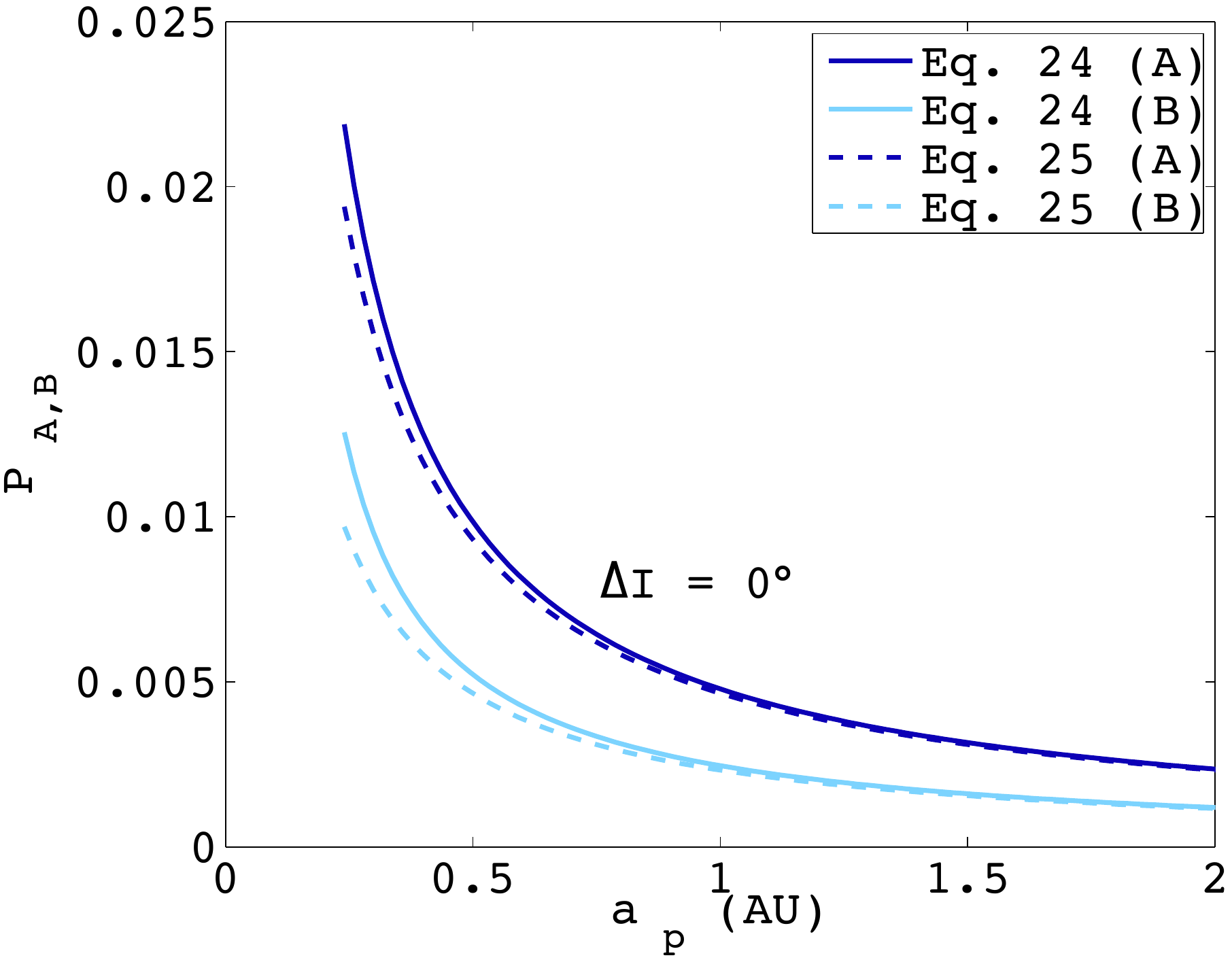}  
		\label{fig:Prob_vs_ap_2}  
	\end{subfigure}
	\caption{In (a) we show the probability of transitability on stars A and B (Eq.~\ref{eq:prob_transitability}) as a function of $a_{\rm p}$, for three different mutual inclinations. The horizontal dashed lines are calculated using the first-order approximate probability (Eq.~\ref{eq:prob_transitability_first_order}). In (b) we zoom in on the $\Delta I = 0^{\circ}$ case. As a comparison, we show the equivalent single star probability in dashed lines (Eq.~\ref{eq:single_star_prob}).
}\label{fig:Prob_vs_ap}  
\end{center}  
\end{figure*}

In Fig.~\ref{fig:Prob_vs_ap_1} we demonstrate Eq.~\ref{eq:prob_transitability} on an example circumbinary system, comprised of a binary with $M_{\rm A}=1M_{\odot}$, $R_{\rm B} = 1R_{\odot}$, $M_{\rm B} = 0.5M_{\odot}$, $R_{\rm B} = 0.5 R_{\odot}$ and $a_{\rm bin} = 0.082$ AU ($T_{\rm bin}=7$ days). The planet semi-major axis is varied from 0.24 AU to 2 AU. The three mutual inclinations are $0^{\circ}$, $5^{\circ}$ and $10^{\circ}$. 

As $\Delta I$ is increased the probability of transitability is increased significantly. For the misaligned cases, $P_{\rm B} > P_{\rm A}$. In Fig.~\ref{fig:Prob_vs_ap_2} we zoom in on the coplanar case. As a comparison, we show the transit probability on a single star of radius $R_{A,B}$, calculated using

\begin{equation}
\label{eq:single_star_prob}
P_{\rm A,B} = \frac{R_{\rm A,B}}{a_{\rm p}}.
\end{equation}

For coplanar systems the probability of transitability reduces to

\begin{equation}
\label{eq:prob_transitability_coplanar}
P_{\rm A,B} = \frac{R_{\rm A,B}}{a_{\rm p} - a_{\rm A,B}},
\end{equation}
which comes from setting $\Delta I=0$ in Eq.~\ref{eq:prob_transitability} and using a small angle approximation to remove the $\sin$ function. This equation matches \citet{welsh12}, who derived an analytic estimate for the transit probability\footnote{In fact, their derivation was for the probability of transitability, despite not using that name.} under the assumption of $\Delta \Omega=0$ and static orbits, although it was duly noted that circumbinary orbits precess. Equation~\ref{eq:prob_transitability_coplanar} is close to the single star probability but slightly higher, particularly at short periods. This is because the stars are brought closer to the planets by their orbital motion.

The crucial difference to the single star case is that Eq.~\ref{eq:single_star_prob} decreases towards zero for large semi-major axes, but for circumbinary systems the limit of Eq.~\ref{eq:prob_transitability} is

\begin{equation}
\label{eq:prob_transitability_approx}
\lim_{a_{\rm p} \rightarrow \infty} P_{\rm A,B}= \sin\Delta I,
\end{equation}
which is equal to the first-order derivation in Eq.~\ref{eq:prob_transitability_first_order}. This approximate probability is also applicable to systems with a large $\Delta I$ because the angle  $\zeta$, which encompasses the period-dependence, becomes relatively small.

\begin{figure}  
\begin{center}  
\includegraphics[width=0.5\textwidth]{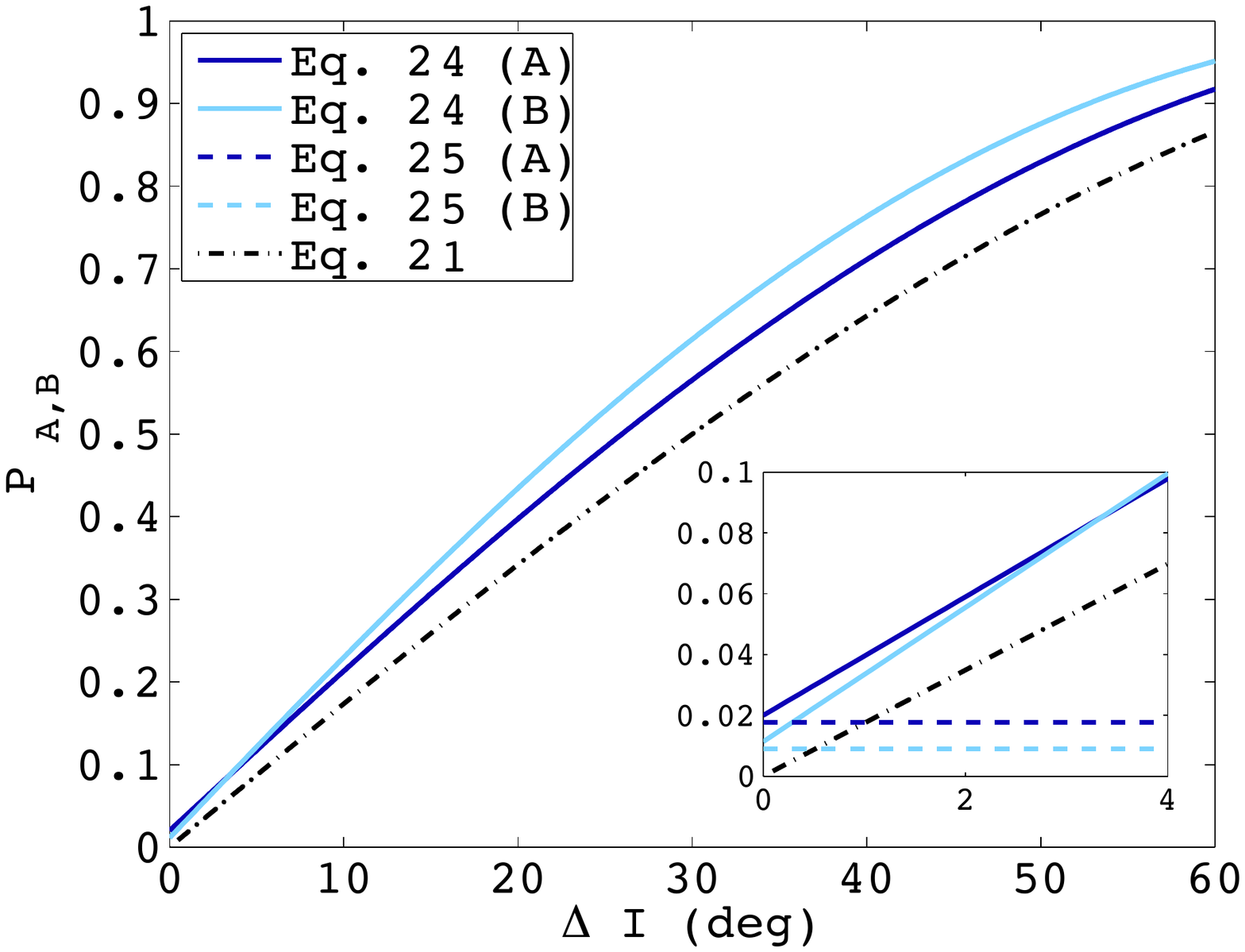}  
\caption{The probability of transitability on stars A and B calculated using Eq.~\ref{eq:prob_transitability}, and the approximation using Eq.~\ref{eq:prob_transitability_first_order}. The bottom right image is zoomed into small mutual inclinations. In this plot the horizontal lines are the equivalent single star transit probabilities (Eq.~\ref{eq:single_star_prob}).}

\label{fig:Prob_vs_MI}
\end{center}  
\end{figure}

In Fig.~\ref{fig:Prob_vs_MI} we demonstrate how the probability of transitability varies with $\Delta I$, using the same circumbinary system as in Fig.~\ref{fig:Prob_vs_ap_1}, but fixing $a_{\rm p} = 0.26$ AU. In the bottom right of this figure we zoom in near $\Delta I = 0^{\circ}$. The curve for the secondary star is seen to overtake that of the primary at around $\Delta I = 3^{\circ}$.

\section{Consequences for eclipsing binaries}
\label{sec:EBs}


Eclipsing binaries are only a small fraction of the total binary population but the easiest binaries to detect photometrically. It was suggested by \citet{borucki84} that they are favourable targets for transit surveys because they positively bias the planetary orbit towards being aligned with the line of sight. In this section we derive the probability of transitability, $P_{\rm A,B}$, under assumption that the binary is known to eclipse. In doing so we quantify what was first noted in Sect.~\ref{sec:criterion} upon deriving Eq.~\ref{eq:TransitabilityInequalityFirstOrder}: the criterion for transitability is easiest to fulfil in the case of eclipsing binaries. The criterion for an eclipse is

\begin{equation}
\label{eq:EclipseCriterion}
\sin\left|\frac{\pi}{2}-I_{\rm bin}\right| \leq \frac{R_{\rm A} + R_{\rm B} - 2\alpha R_{\rm B}}{a_{\rm bin}},
\end{equation}
where $\alpha$ determines whether the criterion is for grazing eclipses ($\alpha = 0$), full eclipses ($\alpha = 1$) or anything in between. Since eclipses occur when $I_{\rm bin} \approx \pi/2$, we can apply the small angle approximation in Eq.~\ref{eq:EclipseCriterion} to deduce that the distribution of $I_{\rm bin}$ for eclipsing binaries is uniform between 

\begin{equation}
\label{eq:Ibin_dist_bounds}
\frac{\pi}{2} \pm \frac{\delta}{a_{\rm bin}},
\end{equation}
where to simplify the equation we have defined
\begin{equation}
\delta = R_{\rm A} + R_{\rm B} - 2\alpha R_{\rm B}.
\end{equation}

Knowing that the binary eclipses with this uniform random distribution of $I_{\rm bin}$, the probability of transitability is the fraction of eclipsing binaries with $I_{\rm bin}$ such that the inequality in Eq.~\ref{eq:beta_factor_inequality} is satisfied. First, multiply Eq.~\ref{eq:EclipseCriterion} by $P_{\rm A,B}$ and insert it into Eq.~\ref{eq:beta_factor_inequality}, using $\beta_{\rm A,B}$ from Eq.~\ref{eq:beta_A}, to obtain

\begin{equation}
\label{eq:EB_calc_1}
\Delta I = P_{\rm A,B}\frac{\delta}{a_{\rm bin}} - \sin^{-1}\left(P_{\rm A,B} \mu_{\rm B,A} \frac{a_{\rm bin}}{a_{\rm p}} \frac{\delta}{a_{\rm bin}} - \frac{R_{\rm A,B}}{a_{\rm p}}\right).
\end{equation}
The inequality from Eq.~\ref{eq:beta_factor_inequality} has disappeared since we are calculating $P_{\rm A,B}$ for a given $\Delta I$. Use the small angle approximation and rearrange to obtain

\begin{equation}
\label{eq:EB_calc_2}
\Delta I = P_{\rm A,B}\delta\left(\frac{1}{a_{\rm bin}} - \mu_{\rm B,A} \frac{1}{a_{\rm p}}\right) - \frac{R_{\rm A,B}}{a_{\rm p}}.
\end{equation}
By solving for $P_{\rm A,B}$ we get
\begin{equation}
\label{eq:EB_calc_3}
P_{\rm A,B} = \twopartdef {\dfrac{\Delta I + \dfrac{R_{\rm A,B}}{a_{\rm p}}}{\delta\left(\dfrac{1}{a_{\rm bin}}-\mu_{\rm B,A}\dfrac{1}{a_{\rm p}}\right)}} {\Delta I < \Delta I_{\rm lim}} {1} {\Delta I \geq \Delta I_{\rm lim}},
\end{equation}where we define
\begin{equation}
\label{eq:MI_lim}
\Delta I_{\rm lim} = \delta\left(\frac{1}{a_{\rm bin}} - \mu_{\rm B,A} \frac{1}{a_{\rm p}}\right) - \frac{R_{\rm A,B}}{a_{\rm p}},
\end{equation}
in order to truncate $P_{\rm A,B}$ at 1. As an example, for a binary with solar and half-solar masses and radii and $a_{\rm bin}=0.3$ AU orbited by a planet at $a_{\rm p}=1.0$ AU, $\alpha=0.5$ and coplanar orbits, Eq.~\ref{eq:EB_calc_3} yields $P_{\rm A} = 0.33$ and $P_{\rm B} = 0.19$. Coplanar orbits correspond to a minimum value of $P_{\rm A,B}$. A slight increase in $\Delta I$ to $0.5^{\circ}$ raises these probabilities to 0.96 and 0.89, respectively. For $\Delta I = 1^{\circ}$ the probability on both stars is 1. The circumbinary geometry is evidently very favourable for transitability on eclipsing binaries. We note for reference that the mean mutual inclination in the transiting circumbinary planets found so far is $1.73^{\circ}$ (see Table~\ref{tab:KeplerDiscoveries}) and that the Solar System mutual inclination distribution roughly follows a $1^{\circ}$ Rayleigh profile relative to the invariant plane (e.g. \citealt{clemence55,lissauer11})

\begin{figure}  
\begin{center}  
\includegraphics[width=0.5\textwidth]{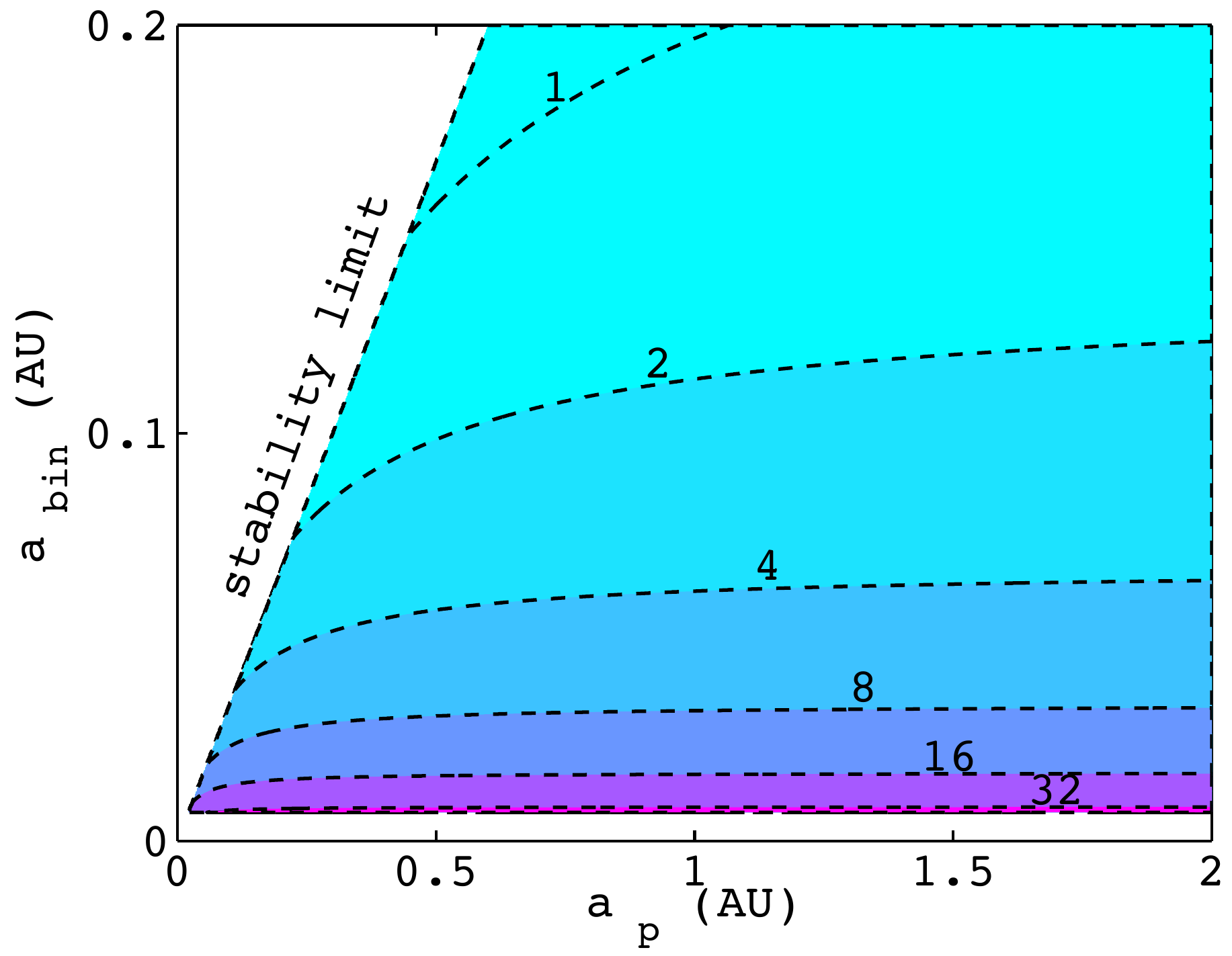}  
\caption{3D histogram of the minimum mutual inclination needed in degrees to guarantee transits on EBs of any orientation, at different binary and planet semi-major axes (Eq.~\ref{eq:MI_lim}).}

\label{fig:MI_lim}
\end{center}  
\end{figure}

For $\Delta I > \Delta I_{\rm lim}$ two things occur: 1) transits are guaranteed on eclipsing binaries of any orientation and 2) transits become possible on non-eclipsing binaries. In Fig.~\ref{fig:MI_lim} we plot $\Delta I_{\rm lim}$ as a function of $a_{\rm bin}$ and $a_{\rm p}$, for a binary with solar and half-solar mass and radius and $a_{\rm bin}$ between 0.007 and 0.2 AU, where the lower limit corresponds to a contact binary: $a_{\rm bin}=R_{\rm A} + R_{\rm B}$. For the eclipse criterion we used $\alpha = 0.5$. The white empty space on the left is the unstable region according to Eq.~\ref{eq:stability_criterion}.

The binary semi-major axis is the biggest factor in the calculation. For the systems in Fig.~\ref{fig:MI_lim} where circumbinary planets have been found so far ($a_{\rm bin} > 0.08$ AU), $\Delta I_{\rm lim}$ is less than $3^{\circ}$.

For closer binaries $\Delta I_{\rm lim}$ rises sharply, reaching a maximum of 38$^{\circ}$ for a contact binary. Transits on very short-period eclipsing binaries are of course possible but Eq.~\ref{eq:MI_lim} shows that not all such binaries can be transited unless there is significant misalignment. In Sect.~\ref{sec:dearth} we apply this work to the {\it Kepler} discoveries so far.

There are similarities between transits on eclipsing binaries and studies of multi-transiting systems orbiting single stars (e.g. \citealt{ragozzine10,gillon11}). Geometrically, one is more likely to find planets transiting a single star where another transiting planet has already been found, compared to around a random star. This is because the mutual inclination distribution of multi-planet systems is not isotropic but weighted towards coplanarity\footnote{\citet{ragozzine10} found it resembles a Rayleigh distribution. The distribution of $\Delta I$ in circumbinary systems is presently unknown, because the detections so far have been highly biased towards coplanarity \citep{martin14}.}. This is analogous to a circumbinary system, if one considers the secondary star as the ``inner planet". There is, however, a fundamental difference between single and binary stars: planets orbiting single stars do so on effectively static orbits. 

\section{Connecting transitability to transits}
\label{sec:connection}

\subsection{Does transitability guarantee transits?}
\label{sec:TransitabilityToTransits}

Transitability alone is not detectable via photometry, one requires an actual transit. A fundamental element of the definition of transitability is that transits are {\it possible but not guaranteed} on any given passing of the binary orbit. By having a transit probability between 0 and 1 for each passing, it is intuitive to think that a transit will eventually happen if observed continuously for a sufficiently long time. This conclusion was shared by \citet{schneider94,welsh12,kratter13,martin14}.

We tested this hypothesis by numerically simulating circumbinary systems over 50 precession periods and looking for transits in cases where transitability occurred.  This was first done for completely random systems taken from the tests in Appendix~\ref{app:test_criterion}. The details are provided in Appendix~\ref{app:test_transitability_to_transits}. Less than 0.3\% of systems managed to evade transit. All of these exceptional cases corresponded to the limit of transitability, where the planet only spends a very short time in transitability, and hence the chance of transiting on any given orbit is small. Transits are expected to occur eventually, but after a time longer than what was simulated.

Second, we constructed systems with 4:1 and 5:1 period commensurabilities, specifically designed to make planets permanently evade transit. The evasion percentage increased slightly but remained less than 1\%. It is probable that this value would eventually drop to zero, but longer simulations would be required. The planets inevitably transit because exact period commensurabilities are not sustainable, owing to perturbations from the binary on the planetary orbit (Sect.~\ref{sec:additional_effects}). Aside for HD 202206 which has a period ratio near 5:1 \citep{correia05}, period commensurabilities have not been observed. This system may not be representative, since it straddles the border between a circumbinary and a two-planet system; the secondary ``star" has a minimum mass of $15 M_{\rm Jup}$. It has also been theorised by \citet{kley14} that circumbinary planets should form between integer period ratios, not in them.

Whilst not an exhaustive proof, our tests indicate that in the vast majority of cases, transitability indeed leads to transit, albeit at an unspecified point in time. 


\subsection{Transits over time}
\label{sec:TransitsOverTime}

The probability of transitability is equivalent to the probability of transit, granted the observer has infinite time. Unfortunately, due to limitations in technology, funding and human life-expectancy, one must strive to capture transits within a finite time. We calculate some example observation times needed in order to observe a transit.

In the case of a single star, after continuous observations of a time equal to $T_{\rm p}$, the planet either will or will not have transited. This is not the case for circumbinary planets, for two reasons:
 
 \begin{itemize}

\item The planet may currently be outside of transitability, but will precess into transitability at a later time.

\item The planet may currently be inside transitability, but the conjunction required for a transit has not yet occurred.

\end{itemize}

The fraction of circumbinary planets transiting therefore increases with time, up to a value specified by Eq.~\ref{eq:prob_transitability}. It is important to know how long an observer must wait to see a transit. It is a strong function of the precession period, since that determines how spaced apart the regions of transitability are. 

An analytic calculation of the time-dependent transit probability is outside the scope of this paper, and has been previously labelled impossible \citep{schneider90}. We instead use numerical N-body simulations.

\begin{figure}  
\begin{center}  
\includegraphics[width=0.5\textwidth]{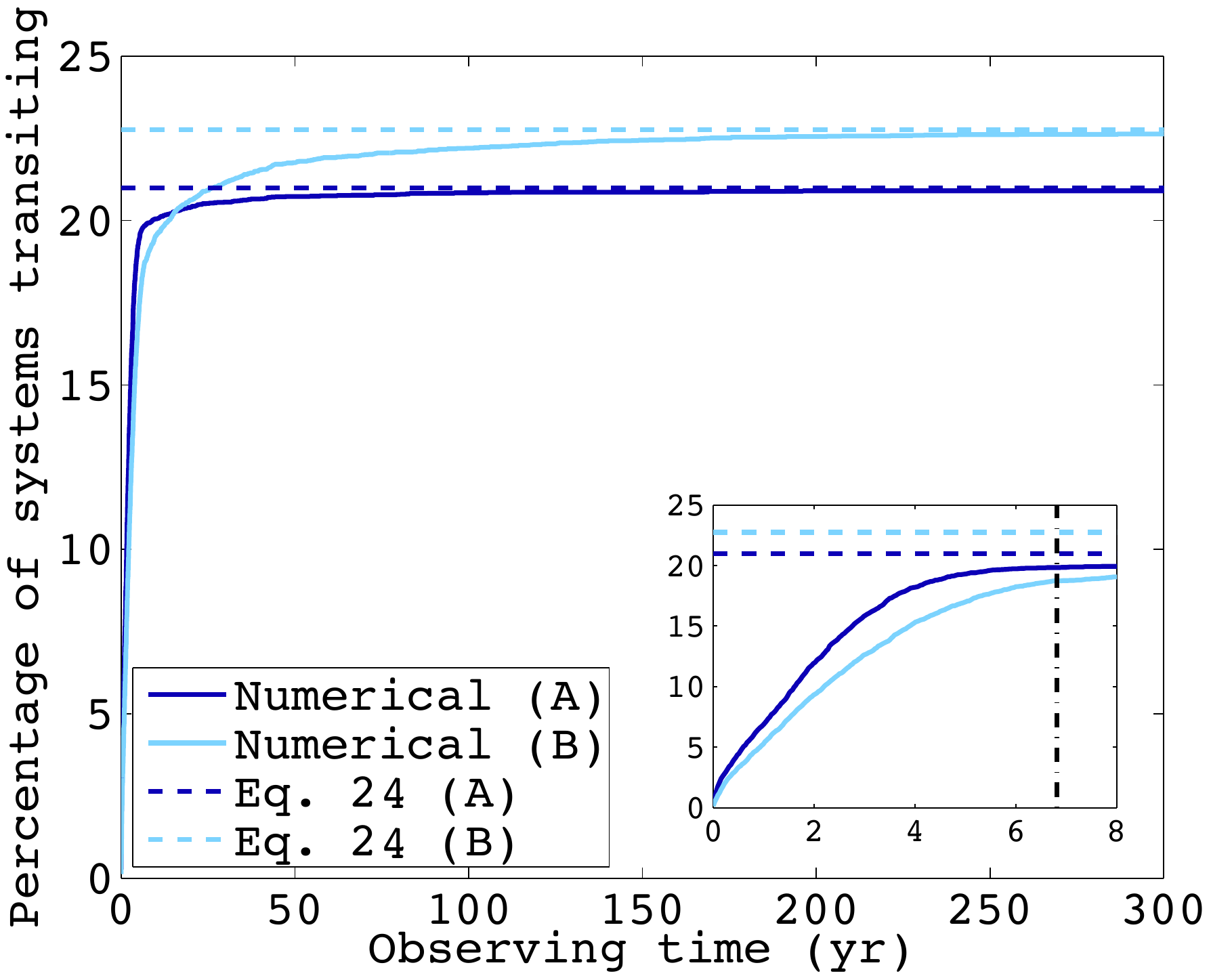}  
\caption{The percentage of systems seen transiting as a function of time from numerical simulations of 10,000 circumbinary systems, and the analytic predictions (Eq.~\ref{eq:prob_transitability}). Systems are counted as transiting after the detection of a single transit. The zoomed figure in the bottom right corner shows the percentage of transiting systems over {\it Kepler}-like observing times. The black vertical line denotes the precession period.}

\label{fig:AsymptoteTest1}
\end{center}  
\end{figure}

In Fig.~\ref{fig:AsymptoteTest1} we demonstrate the percentage of systems seen transiting stars A and/or B as a function of time, using 10,000 simulated circumbinary systems. The primary and secondary stars are solar and half-solar in mass and radius, $T_{\rm bin}=7$ days, $T_{\rm p}=40$ days and $\Delta I=10$. Over time, the percentage of transiting systems reaches the value predicted by Eq.~\ref{eq:prob_transitability}, in agreement with the conclusions of Sect.~\ref{sec:TransitabilityToTransits}. Most of the transiting systems have done so within a single precession period (here $\sim$ seven years).

As an extended test, we took the systems found transiting in in Sect.~\ref{sec:TransitabilityToTransits} and calculated the time taken for primary and secondary transits to occur. The results are provided in Table~\ref{tab:median_wait_time}. Whilst a larger mutual inclination leads to more planets transiting, the median wait time is increased. Generally, a significant number of systems are found transiting within {\it Kepler}-like mission times.

\section{Discussion and applications}
\label{sec:discussion}

\subsection{The circumbinary planets discovered so far}

\begin{table*}
\caption{Probabilities of transit for the circumbinary planets detected so far by {\it Kepler}.} 
\centering 
\begin{tabular}{ ccccccccccccccc } 
\hline\hline 
Name & $M_{\rm A}$ & $M_{\rm B}$ & $R_{\rm A}$ & $R_{\rm B}$ & $a_{\rm bin}$ & $a_{\rm p}$ & $\Delta I$ &\multicolumn{2}{|c}{$P_{\rm A,B}$ \% (all)} &\multicolumn{2}{|c}{$P_{\rm A,B}$ \% (EBs)} &\multicolumn{2}{|c}{$P_{\rm A,B}$ \% (single)}\\ 
[0.5ex] 
& ($M_{\odot}$) & ($M_{\odot}$) & ($R_{\odot}$) & ($R_{\odot}$) & (AU) & (AU) & (deg) & A & B & A & B & A & B \\
\hline
Kepler-16 & 0.69 & 0.20 & 0.65 & 0.23 & 0.22 & 0.71 & 0.31 & 1.04 & 0.91 & 75.5 & 66.1 & 0.42 & 0.15 \\
Kepler-34 & 1.05 & 1.02 & 1.16 & 0.19 & 0.23 & 1.09 & 1.86 & 4.18 & 4.16 & 100 & 100 & 0.50 & 0.47 \\
Kepler-35 & 0.89 & 0.81 & 1.03 & 0.79 & 0.18 & 0.60 & 1.07 & 3.11 & 2.94 & 100 & 100 & 0.78 & 0.61 \\
Kepler-38 & 0.95 & 0.27 & 1.78 & 0.27 & 0.15 & 0.46 & 0.18 & 2.28 & 0.79 & 41.2 & 14.4 & 1.80 & 0.27 \\
Kepler-47b & 1.04 & 0.46 & 0.84 & 0.36 & 0.08 & 0.30 & 0.27 & 1.93 & 1.26 & 39.5 & 25.8 & 1.30 & 0.56 \\
Kepler-47c & 1.04 & 0.46 & 0.84 & 0.36 & 0.08 & 0.99 & 1.16 & 2.48 & 2.32 & 50.8 & 47.6 & 0.39 & 0.17 \\
Kepler-64 & 1.50 & 0.40 & 1.75 & 0.42 & 0.18 & 0.65 & 2.81 & 6.54 & 6.66 & 100 & 100 & 1.25 & 0.30 \\
Kepler-413 & 0.82 & 0.52 & 0.78 & 0.48 & 0.10 & 0.36 & 4.02 & 8.98 & 9.18 & 100 & 100 & 1.01 & 0.62 \\
KIC9632895 & 0.93 & 0.19 & 0.83 & 0.21 & 0.18 & 0.93 & 2.30 & 4.68 & 5.09 & 100 & 100 & 0.49 & 0.12 \\
\hline 
\multicolumn{14}{l}{\footnotesize{{\bf Refs:} \citet{doyle11,welsh12,orosz12a,orosz12b,schwamb13,kostov13,kostov14}}}\\
\multicolumn{14}{l}{\footnotesize{\citet{welsh14}}}\\
\multicolumn{14}{l}{\footnotesize{{\bf Note:} Kepler-47d is excluded because it has not yet been published and lacks a value for $\Delta I$.}}\\

\end{tabular}
\label{tab:KeplerDiscoveries}
\end{table*}

Our first application is to calculate the transitability probabilities for the {\it Kepler} discoveries so far, assuming of course that we do not have a priori knowledge of transits and eclipses\footnote{Otherwise you would have a boring table full of 100\%'s.}. In Table.~\ref{tab:KeplerDiscoveries} we calculate the probability of transitability on binaries of any orientation (Eq.~\ref{eq:prob_transitability}) and on eclipsing binaries (Eq.~\ref{eq:EB_calc_3}), where for the latter we used $\alpha=0.5$ to define eclipses. The equivalent single star probability was calculated using Eq.~\ref{eq:single_star_prob}. In the table we include all necessary variables for the calculations. In more than half of the cases, transits are guaranteed on eclipsing binaries of any orientation. 

\subsection{Multi-planet circumbinary systems}

Only one multi-planet circumbinary system has been discovered so far (Kepler-47, \citealt{orosz12b}). \citet{kratter13} considered an eclipsing binary with a known transiting planet, and calculated the likelihood of a second planet being seen transiting. They derived an analytic probability for whether or not the binary and planet orbits would overlap on the sky, under the assumption that the binary is perfectly edge-on ($I_{\rm bin}=\pi/2$). In fact, what they calculated was the probability of transitability. Their derivation does not include precession, and consequently underestimates the probability. 

Based on the work in Sect.~\ref{sec:EBs}, any additional planets with $\Delta I$ greater than the first transiting planet are guaranteed to enter transitability at some point.

\subsection{{\it Kepler's} eclipsing binary catalog}

The {\it Kepler} telescope, with its four years of continuous observations and exquisite precision, has provided the most comprehensive catalog of EBs to date \citep{slawson11}. We used the online beta version of this catalog\footnote{http://keplerebs.villanova.edu/ maintained by Andres Prsa et al.} to test our transitability criterion on hypothetical orbiting planets. From the catalog we obtained $M_1,M_2,R_1,R_2$ and $a_{\rm bin}$, which were derived from stellar temperatures calculated in \citet{armstrong13b} using a method explained in \citet{armstrong14}. Only systems with a morphology parameter less than 0.5 were used, corresponding to detached EBs (see \citealt{matijevic12} for details). The binary inclination was randomised between the bounds defined in Eq.~\ref{eq:Ibin_dist_bounds}\footnote{It is possible to obtain a true value of $I_{\rm bin}$, however the only published version is in the now-outdated catalog of \citet{slawson11}, and contains errors.}.

\begin{figure}  
\begin{center}  
\includegraphics[width=0.5\textwidth]{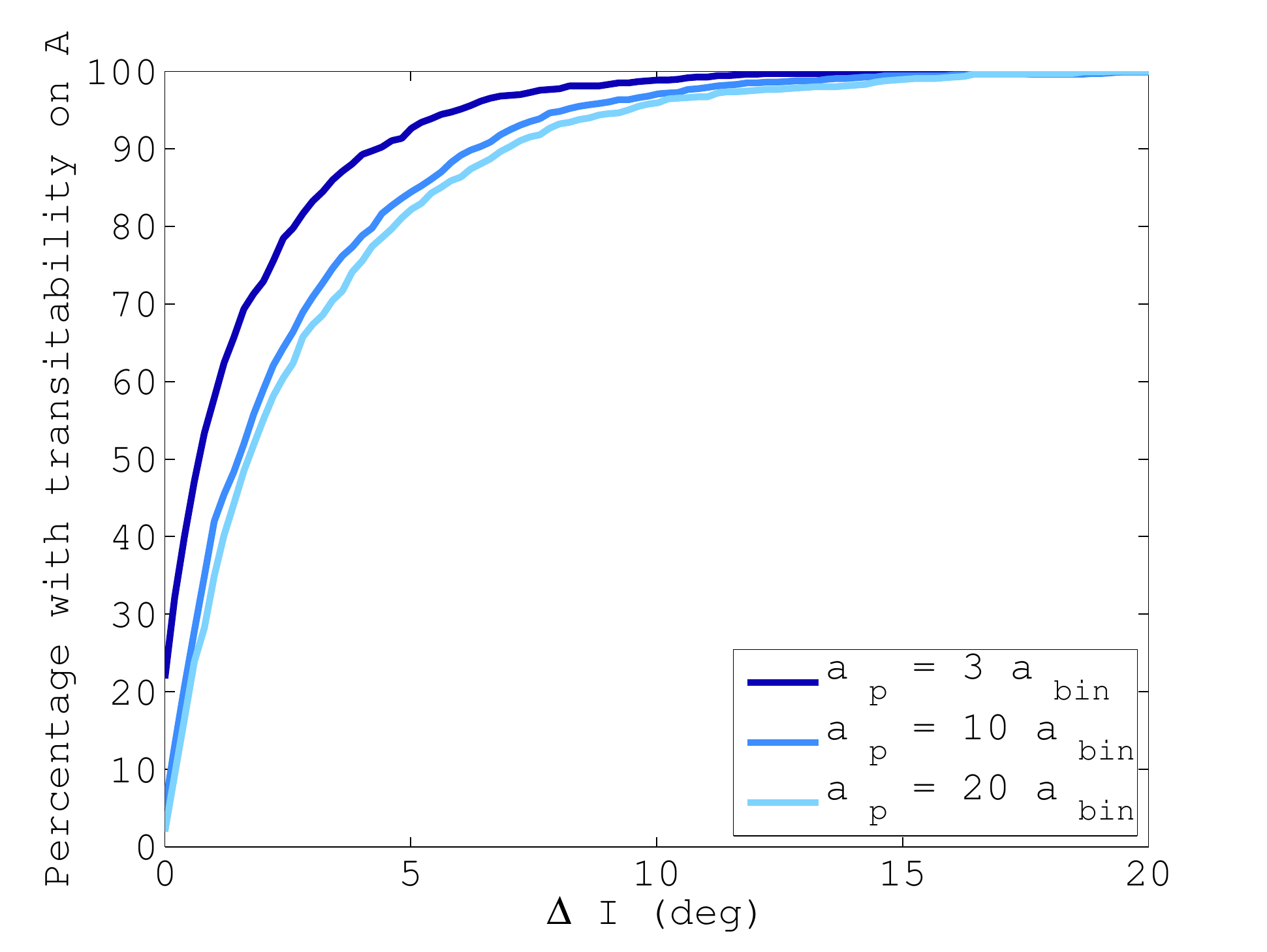}  
\caption{The percentage of EBs found by {\it Kepler} on which there would be transitability by a putative planet with different values of $\Delta I$ and $a_{\rm p}/a_{\rm bin}$.}

\label{fig:KeplerTest_1}
\end{center}  
\end{figure}

The remaining quantities needed for Eq.~\ref{eq:beta_factor_inequality} are $\Delta I$ and $a_{\rm p}$. Given the distribution of circumbinary planets is presently poorly known and subject to strong biases, we considered a wide range of potential values. In Fig.~\ref{fig:KeplerTest_1} we calculated the percentage of {\it Kepler} EBs on which there would be transitability by a putative planet, with $\Delta I$ between $0^{\circ}$ and $10^{\circ}$ and $a_{\rm p}/a_{\rm bin} = 3, 10, 20$. The results are only shown for the primary star, since the plot for the secondary star is indistinguishable.

Consistent with earlier sections, the biggest factor is the mutual inclination, with a higher $\Delta I$ leading to a greater chance of transitability. Transitability is favoured for smaller values of $a_{\rm p}$, but this dependency diminishes at larger mutual inclinations.

The {\it Kepler} EB catalog would benefit from extended photometric observations by the future {\it PLATO} telescope \citep{rauer13}, in order to find new circumbinary planets that have moved into transitability during the $\sim$ eight years between missions. This may include additional planets in known circumbinary systems.

\subsection{On the dearth of planets around short-period binaries}
\label{sec:dearth}

An observed trend has been the lack of circumbinary planets around the closest binaries; the shortest-period binary hosting a planet is Kepler-47 with $T_{\rm bin} = 7.4$ days \citep{orosz12b}. This is despite the median\footnote{In the published version of the paper we accidentally wrote that this was the ``mean" period.} period of the EB catalog being 2.8 days. This raises various questions about the ability to form planets in such an environment, particularly in the presence of tertiary stellar companion, as is often the case for very tight binaries according to theory \citep{mazeh79,fabrycky07} and observations \citep{tokovinin06}.

\begin{figure}  
\begin{center}  
\includegraphics[width=0.5\textwidth]{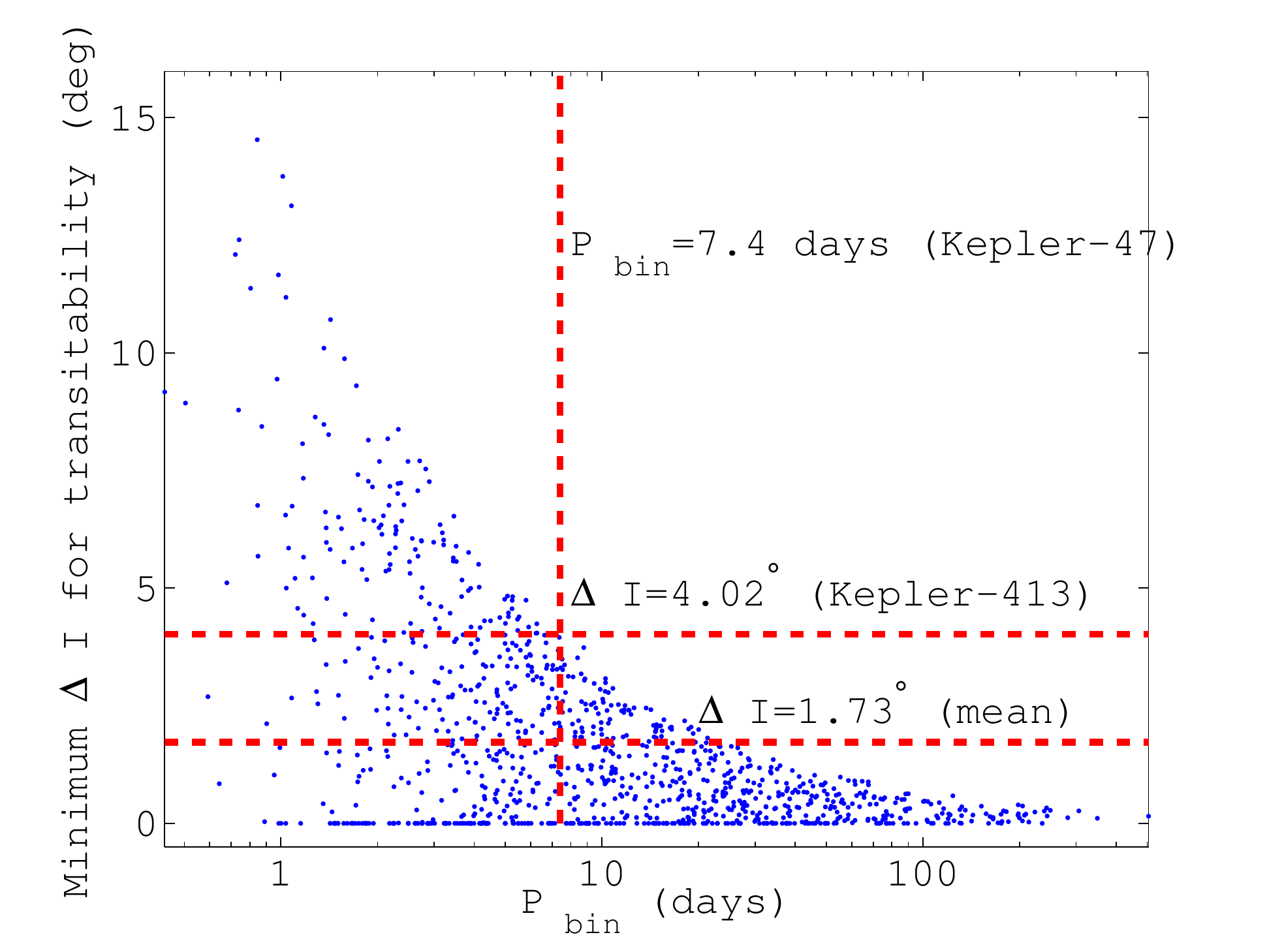}  
\caption{The minimum mutual inclination needed for transitability on each of the EBs found by {\it Kepler}.}

\label{fig:KeplerTest_2}
\end{center}  
\end{figure}

The reason why EBs are preferentially found at short periods despite a smaller natural occurrence \citep{tokovinin06} is because there is a greater range of $I_{\rm bin}$ that allow for an eclipse (Eq.~\ref{eq:Ibin_dist_bounds}). When the EB is highly inclined, however, the planet itself needs a greater misalignment in order for transitability to occur. We demonstrate this in Fig.~\ref{fig:KeplerTest_2}, where we calculate the minimum $\Delta I$ needed to see transitability on the primary star of each EB in the {\it Kepler} catalog, taking $a_{\rm p}/a_{\rm bin} = 3.5$. As a reference, we show the mean and maximum mutual inclinations from the {\it Kepler} discoveries so far, although these are biased towards being small \citep{martin14}. 

For $T_{\rm bin}<7.4$ days, a misalignment of $1.73^{\circ}$ results in 60\% of planets missing transitability. For those systems misaligned enough for transitability on the shortest period binaries, there should be transits within the {\it Kepler} time series, for two reasons: 1) the precession period is only a couple of years long\footnote{If we assume the observed over density of planets at $T_{\rm p} \sim 5T_{\rm bin}$ extends to very close binaries.}, so the planet and binary orbits would have intersected at least once during the {\it Kepler} mission and 2) these are very tight systems, so we are likely to have observed one or probably more transits whilst in transitability.

The dearth of planets may also be explained by stellar noise in the light curves. Binaries with periods this short are expected to be tidally locked, which leads to faster rotation and increased star spots, which may inhibit detections.

The current null detection likely remains significant for planets that are misaligned by at least a few degrees, but there may remain some coplanar ones that are undetectable by {\it Kepler}. Transits on contact binaries require an even higher level of mutual inclination. Discoveries around contact binaries may also be hindered by {\it Kepler's} 30-minute cadence, which is potentially too long to adequately sample its orbit in the search for transits.


\subsection{Limitations}
\label{sec:limitations}

\subsubsection{Eccentric Systems}
\label{sec:eccentric_systems}

The addition of eccentricity, to both the binary and planet orbits, introduces two complexities. First, the geometry is complicated since we lose circular symmetry, and there are two additional angles to consider: $\omega_{\rm bin}$ and $\omega_{\rm p}$ (the arguments of periapse). Furthermore, the orbital dynamics cause $\omega_{\rm p}$ to be time-dependent, further complicating the situation.

\begin{figure}  
\begin{center}  
\includegraphics[width=0.5\textwidth]{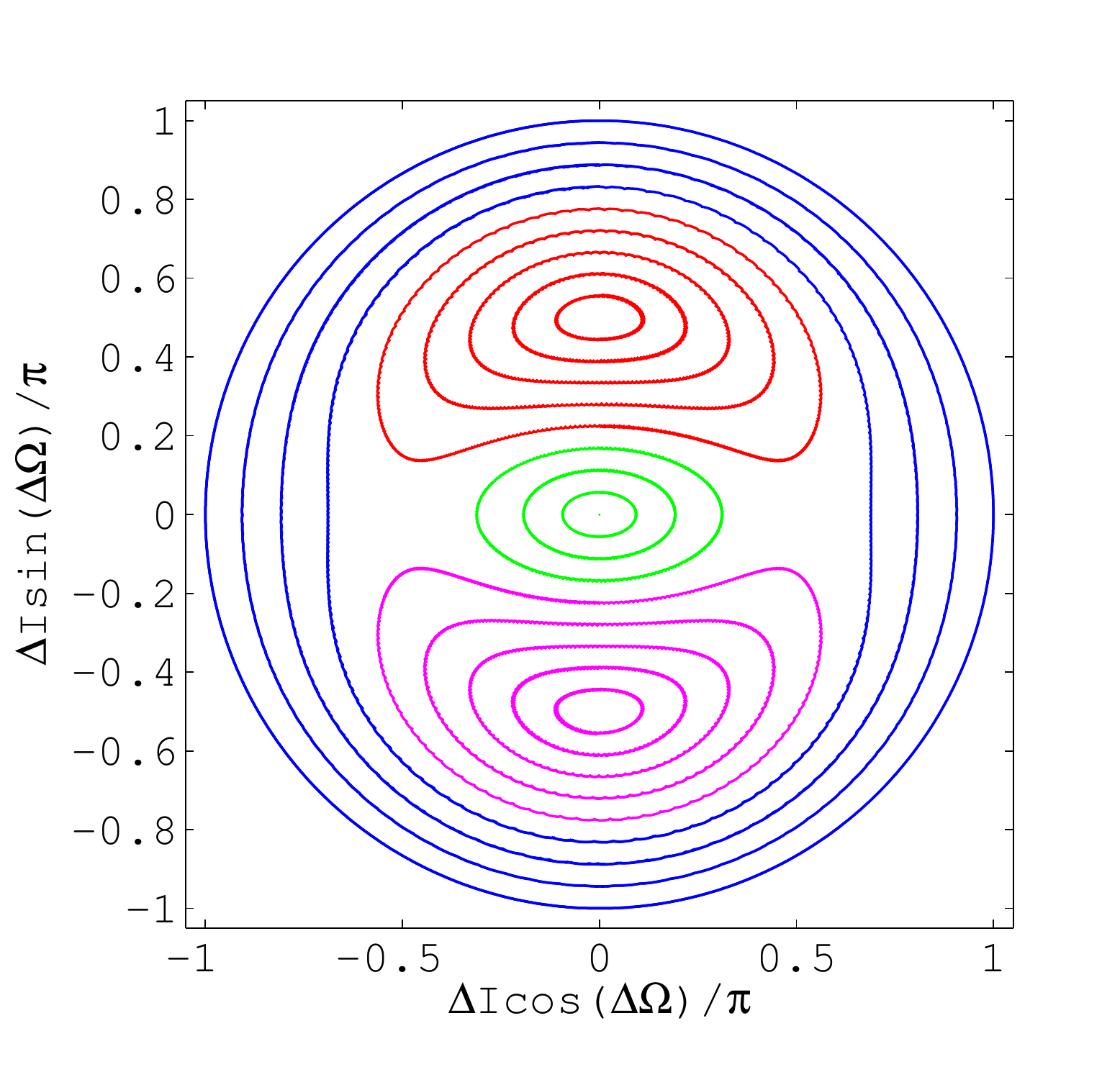}  
\caption{The same circumbinary systems as in Fig.~\ref{fig:DoolinBlundellTest} but with $e_{\rm bin}=0.5$.}

\label{fig:DoolinBlundellTest_ecc}
\end{center}  
\end{figure} 

Second, the precession cycle is more complex when the binary is eccentric. In Fig.~\ref{fig:DoolinBlundellTest_ecc} we demonstrate the precession of the same system as in Fig.~\ref{fig:DoolinBlundellTest}, but with $e_{\rm bin} =0.5$. The mutual inclination is no longer constant. There are two islands of libration, centred on $\Delta \Omega=0$ and $\Delta I = \pi/2$ (red) and $\Delta I = -\pi/2$ (magenta), within which $\Delta \Omega$ does not circulate through 0 to $2\pi$. We therefore lose two of the assumptions made in Sect.~\ref{sec:criterion}.

We ran numerical simulations to test the ability of our criterion in Eq.~\ref{eq:beta_factor_inequality} to predict transitability, in the case of eccentric systems. The simulations show that the error is only of order $\sim 2\%$ (Appendix~\ref{app:eccentric_systems}). Furthermore, the results suggest that eccentricity actually makes transitability more likely, although a more detailed study is needed to confirm this.

\subsubsection{Additional dynamical effects}
\label{sec:additional_effects}

A circumbinary system is a three-body problem and hence not solvable analytically. Our analytic treatment of it as a pair of Keplerians plus orbital precession encompasses the majority of the physics, but neglects some smaller amplitude effects.

\begin{figure}  
\begin{center}  
\includegraphics[width=0.5\textwidth]{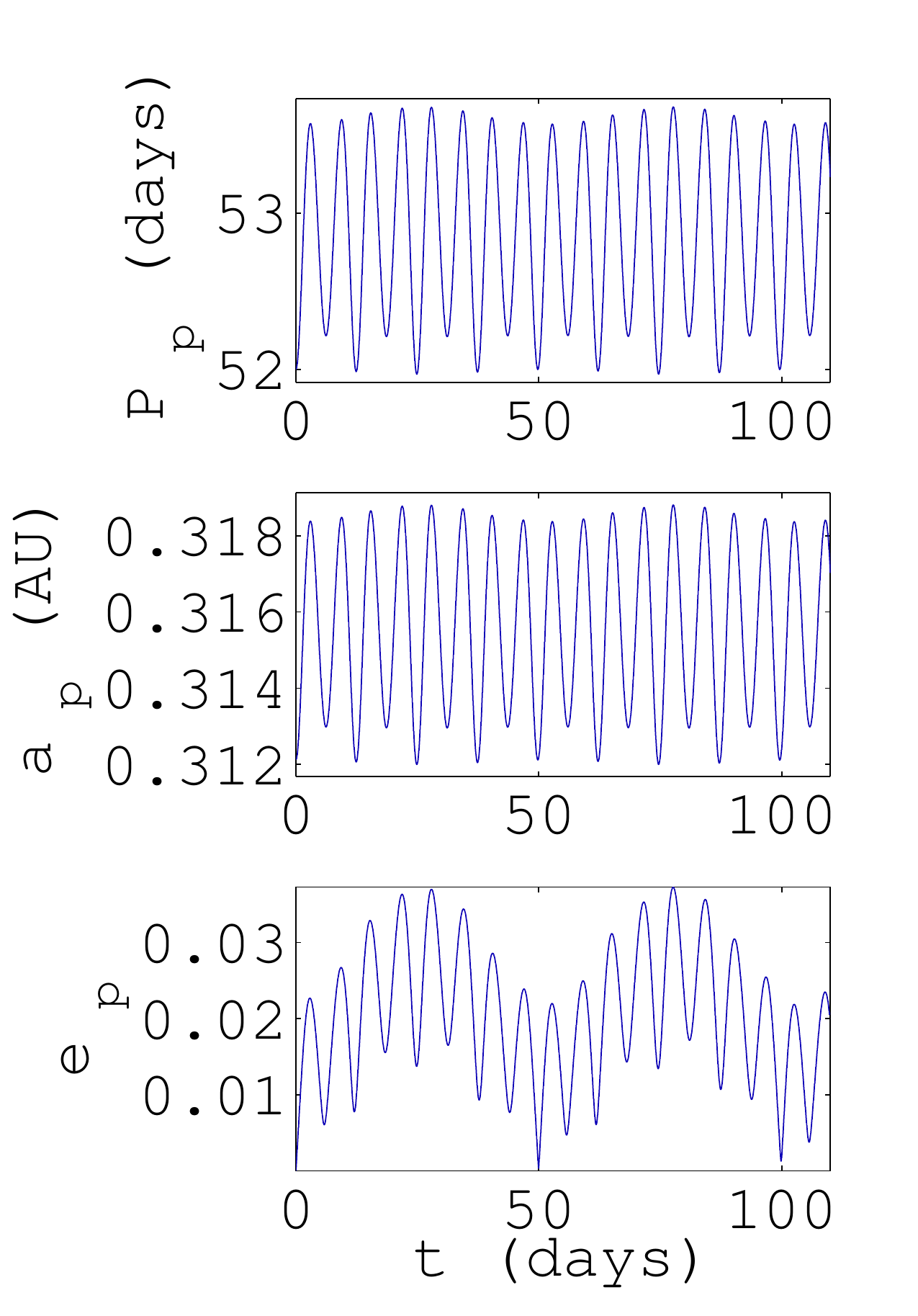}  
\caption{Osculating orbital elements for an example circumbinary system, viewed over a timespan slightly greater than $2P_{\rm p}$.}

\label{fig:other_effects}
\end{center}  
\end{figure}

The semi-major axis and period, which we assumed to be constant, experience slight variations over time because of perturbations from the binary. There are also small variations in the eccentricity. In fact, even a system with an initially circular planetary orbit will obtain some eccentricity over time, and hence the complications of Sect.~\ref{sec:eccentric_systems} are to a certain extent unavoidable. In Fig.~\ref{fig:other_effects} we present an example of the variation in planet period (top), semi-major axis (middle) and eccentricity (bottom) for a binary with solar and half-solar mass and radius stars, $T_{\rm bin} = 10$ days and a planet with an initial period of $T_{\rm p}$ = 52 days and an initially circular orbit. There is a short-term variation with period $1/2T_{\rm bin}$ and a longer-term modulation with period $T_{\rm p}$. A comprehensive analysis of the orbital dynamics of circumbinary systems can be found in \citet{leung13}.

We have also neglected any effects that may be imposed by a tertiary star. Tertiary stars are very commonly found around binaries \citep{tokovinin06} and, if close enough, may affect the precession and observability of any circumbinary planets. It is also possible that tertiary stars with sufficient influence would also have hindered planet formation; this case would consequently be rare. 

Additional dynamical effects may also arise in the presence of multi-planet circumbinary systems, although it remains to be seen if these would have a noticeable impact in comparison with the perturbations from the binary.


\subsubsection{The detectability of transits}

Whilst we have shown that transits are more likely at a higher mutual inclination, we have not considered their detectability. Even in the simplest case of a coplanar planet, there are significant variations in transit timing \citep{armstrong13a} and duration \citep{liu14}, making detections more difficult than in the single star case. For highly misaligned systems, the transits will be aperiodic, and perhaps even singular \citep{martin14}. Also, because of the precession, the transit signature may disappear for a while. It was not until the discovery of the ninth transiting circumbinary planet - the $4.1^{\circ}$ misaligned Kepler-413 by \citet{kostov14} - that the effects of time-dependent transitability were readily apparent. This does demonstrate, however, that detection techniques are improving in order to discover these more complicated systems.

Another unsolved question is the predictability of transits. \citet{martin14} showed that the sequence of primary and secondary transits observed for misaligned systems is highly sensitive to the input orbital parameters, e.g. altering the starting planet inclination by as little as $1^{\circ}$ may completely change the transit number and timing. The small variations in orbital elements described in Sect.~\ref{sec:additional_effects} will have to be accounted for, although the regularity of the variation seen in Fig.~\ref{fig:other_effects} is promising. Predictability is not a concern for blind, continuous surveys like {\it Kepler}, {\it TESS} and {\it PLATO}. It is crucial, however, for any targeted follow-up observations.

\section{Conclusion}
\label{sec:conclusion}

The geometry and orbital dynamics unique to circumbinary planets make them very likely to exhibit transitability, and hence transit at some time. Transits provide exoplanetary science with a wealth of information. This probability can be increased to one when observing eclipsing binaries. Furthermore, there is a relatively weak dependence on period. This allows an extension of transit surveys to the outer regions of exoplanetary systems, with applications including including the atmospheric characterisation of cold exoplanets. The high transit probability bodes well for complementary observations of circumbinary planets found with other techniques, such as ETVs \citep{borkovits11}, radial velocimetry \citep{konacki09} and astrometry \citep{salmann14}. For these reasons, circumbinary planets can be seen as more than just exotic examples of nature's diversity but as practical tools in astronomy.

\section{Acknowledgements}
A big thanks to our mate Dave Armstrong for providing the masses and radii of the {\it Kepler} EBs. We thank Darin Ragozzine for kindly reading through the paper and providing some insightful comments. We are also appreciative of the very useful and interesting conversations on the subject with Rosemary Mardling, St\'ephane Udry and Bill Welsh. Finally, we thank an anonymous referee for taking the time to assess our paper and providing a timely report. D.~V.~Martin is funded by the Swiss National Science Foundation. A.~H.\,M.\,J.~Triaud is a Swiss National Science Foundation fellow under grant number P300P2-147773. Our work could not have happened without the important work and dedication of the {\it Kepler} team who compile updated and publicly available candidate lists and stellar catalogs on the \href{https://archive.stsci.edu/}{MAST} repository and at \href{http://keplerebs.villanova.edu/}{Villanova University}. We also made an extensive use of \href{http://adsabs.harvard.edu}{ADS}, \href{http://arxiv.org/archive/astro-ph}{arXiv} and the two planets encyclopaediae \href{exoplanet.eu}{exoplanet.eu} and \href{exoplanets.org}{exoplanets.org}. We thanks the teams behind these online tools, which greatly simplify our research.

\appendix

\section{Testing the analytic criterion}
\label{app:test_criterion}

The test consisted of creating 10,000 random circumbinary systems, integrating them over an entire precession period\footnote{The integration time was $1.1T_{\rm prec}$, where $T_{\rm prec}$ was calculated using \citet{farago10}, and the factor of 1.1 allows for any errors in their formula.}, and checking if the planet and stellar orbits overlapped at any time. The orbital parameters were drawn from a uniform distribution between the maximum and minimum bounds listed below.

\begin{itemize}

\item For the binary: $M_{\rm A}$ between 0.5 and 1.5 $M_{\odot}$; $M_{\rm B}$ calculated using a random mass ratio $q$ drawn between 0.2 and 1.0; $R_{\rm A}$ and $R_{\rm B}$ calculated using a mass-radius relation of \citet{kippenhahn94}; $T_{\rm bin}$ between 5 and 50 days; $e_{\rm bin} = 0$ and the starting orbital phase between $0$ and $2\pi$.

\item For the planet: $M_{\rm p} = 0$; $R_{\rm p} = 0$; $T_{\rm p} / T_{\rm bin}$ between 4 and 10; $e_{\rm p}=0$ and the starting orbital phase between $0$ and $2\pi$.

\item Two uniform mutual inclination distributions were tested: $0^{\circ}$ to $5^{\circ}$ (test 1) and $0^{\circ}$ to $30^{\circ}$ (test 2). We drew $\Delta \Omega$ between $0$ and $2\pi$. The orientation of the circumbinary system on the sky was randomised using a uniform 3D rotation algorithm by \citet{arvo92}, which creates a uniform distribution of $\cos I_{\rm bin}$.
\end{itemize}

These orbital parameters are not completely arbitrary, since they encompass all of the circumbinary systems discovered to date. The planet period was chosen with respect to the binary period so that the stability limit was respected (Eq.~\ref{eq:stability_criterion}).

The test results are shown in Table~\ref{tab:CriterionTest}. The result for each system was put into one of four categories:
\begin{itemize}

\item Cat. 1: Eq.~\ref{eq:beta_factor_inequality} predicted transitability and the numerical simulation matched this.

\item Cat. 2: Eq.~\ref{eq:beta_factor_inequality} predicted no transitability and the numerical simulation matched this.

\item Cat. 3: Eq.~\ref{eq:beta_factor_inequality} predicted transitability but the numerical simulation did not show transitability

\item Cat. 4: Eq.~\ref{eq:beta_factor_inequality} predicted no transitability but the numerical simulation showed transitability

\end{itemize}

The last two categories correspond to errors in the analytic formula.


\begin{table*}
\caption{Testing the transitability criterion} 
\centering 
\begin{tabular}{l c | c | cc | cc | cc | cc | cc |} 
\hline\hline 
test & $\Delta I_{\rm max}$ (deg) &\multicolumn{2}{|c}{Cat. 1} &\multicolumn{2}{|c}{Cat. 2} & \multicolumn{2}{|c}{Cat. 3} & \multicolumn{2}{|c}{Cat. 4} & \multicolumn{2}{|c}{error $\%$} \\ 
\hline 
& & A & B & A & B & A & B & A & B & A & B\\
1 & 5 & 618 & 643 & 9381 & 9352 & 1 & 2 & 0 & 3 & 0.01 & 0.03\\
2 & 30 & 2884 & 3105 & 7106 & 6884 & 9 & 4 & 1 & 4 & 0.06 & 0.06\\
\hline 

\end{tabular}
\label{tab:CriterionTest}
\end{table*}

\section{Testing the connection between transitability and transits}
\label{app:test_transitability_to_transits}

\begin{table*}
\caption{Testing if transitability leads to transits} 
\centering 
\begin{tabular}{l c | c | c | cc | cc | cc} 
\hline\hline 
test & $\Delta I_{\rm max}$ (deg) & $T_{\rm p} / T_{\rm bin}$ &\multicolumn{2}{|c}{no. in transitability} &\multicolumn{2}{|c}{no. evading transit} &\multicolumn{2}{|c}{evasion \%} \\ 
\hline 
& & & A & B & A & B & A & B\\
1 & 5 & 4 - 10 & 618 & 641 & 0 & 2 & 0 &  0.31\\
2 & 30 & 4 - 10 & 2878 & 3096 & 7 & 9 & 0.24 & 0.29\\
3 & 5 & 4 & 621 & 663 & 3 & 4 & 0.48 & 0.60\\
4 & 5 & 5 & 593 & 662 & 2 & 6 & 0.34 & 0.91\\
5 & 30 & 4 & 3044 & 3370 & 15 & 30 & 0.49 & 0.89\\
6 & 30 & 5 & 2956 & 3240 & 11 & 14 & 0.37 & 0.43\\

\hline 

\end{tabular}
\label{tab:TransitabilityToTransits}
\end{table*}

As a starting point, we took the circumbinary systems from Appendix.~\ref{app:test_criterion} that were found in transitability (test 1: $\Delta I_{\rm max} =  5^{\circ}$ and test 2: $\Delta I_{\rm max} = 30^{\circ}$). The small number of error cases in Appendix.~\ref{app:test_criterion} were avoided. For each system we integrated over 50 precession periods to look for primary and secondary transits. 

Next, we tested whether planets on periods commensurate with that of the binary would manage to permanently evade transit. We created 10,000 systems with the same parameters as in Appendix~\ref{app:test_criterion}, but with $T_{\rm p}/T_{\rm bin}$ fixed at 4 and at 5. This led to four additional tests: 3) $\Delta I_{\rm max} =  5^{\circ}$ and $T_{\rm p}/T_{\rm bin} = 4$; 4) $\Delta I_{\rm max} =  5^{\circ}$ and $T_{\rm p}/T_{\rm bin} = 5$; 5) $\Delta I_{\rm max} =  30^{\circ}$ and $T_{\rm p}/T_{\rm bin}=4$; and 6) $\Delta I_{\rm max} =  30^{\circ}$ and $T_{\rm p}/T_{\rm bin} = 5$.

In Table~\ref{tab:TransitabilityToTransits} we show the number of systems in transitability and the number of systems evading transit during 50 precession periods, for the six tests outlined above.

\section{Median transit wait times}

For each of the transiting circumbinary systems in Appendix~\ref{app:test_transitability_to_transits} we calculated the time of first transit on each star. In Table.~\ref{tab:median_wait_time} we show the median time taken. We also show this time as a fraction of the precession period. In the last two columns we show the percentage that are seen transiting at least once within four years, which comparable to a {\it Kepler}-like mission.

The purposes of these numbers is to provide a rough estimate of the time needed for transits to occur, and to motivate the fact that they can occur within realistic timeframes.

\begin{table*}
\caption{The time taken for primary and secondary transits to occur} 
\centering 
\begin{tabular}{l c | c | c | cc | cc | cc} 
\hline\hline 
test & $\Delta I_{\rm max}$ (deg) & $T_{\rm p} / T_{\rm bin}$ &\multicolumn{2}{|c}{median time (yr)} &\multicolumn{2}{|c}{scaled median time} &\multicolumn{2}{|c}{percentage $<$ 4 yr} \\ 
\hline 
& & & A & B & A & B & A & B\\
1 & 5 & 4 - 10 & 6.6 & 8.4 & 0.24 & 0.31 & 37 & 33\\
2 & 30 & 4 - 10 & 18.5 & 28.0 & 0.50 & 0.74 & 19 & 16\\
3 & 5 & 4 & 2.4 & 3.4 & 0.24 & 0.35 & 68 & 55\\
4 & 5 & 5 & 3.4 & 4.5 & 0.22 & 0.30 & 55 & 46\\
5 & 30 & 4 & 6.9 & 11.7 & 0.63 & 1.03 & 37 & 27\\
6 & 30 & 5 & 10.2 & 16.5 & 0.57 & 0.85 & 29 & 21\\

\hline 

\end{tabular}
\label{tab:median_wait_time}
\end{table*}

\section{Eccentric systems}
\label{app:eccentric_systems}

We ran a single set of numerical simulations, similar to in Appendix~\ref{app:test_criterion}, to test how accurate the transitability criterion was for eccentric systems. All of the parameters used were the same as in Appendix~\ref{app:test_criterion}, except $e_{\rm bin}$ and $e_{\rm p}$ were varied uniformly between 0 and 0.5. The mutual inclination maximum was also increased to $60^{\circ}$, so we could probe the islands of libration seen in Fig.~\ref{fig:DoolinBlundellTest_ecc}. Adding eccentricity can make a system unstable, so we removed any such cases according to either the stability criterion in \citet{holman99}. Since \citet{holman99} includes $e_{\rm bin}$ but not $e_{\rm p}$, we used the periapsis distance $a_{\rm p}(1-e_{\rm p})$ as a approximate means of including the effects of planet eccentricity. Additionally, we monitored and removed planets that were ejected during the numerical simulations. The results are shown in Table.~\ref{tab:EccTest}.

\begin{figure}  
\begin{center}  
\includegraphics[width=0.5\textwidth]{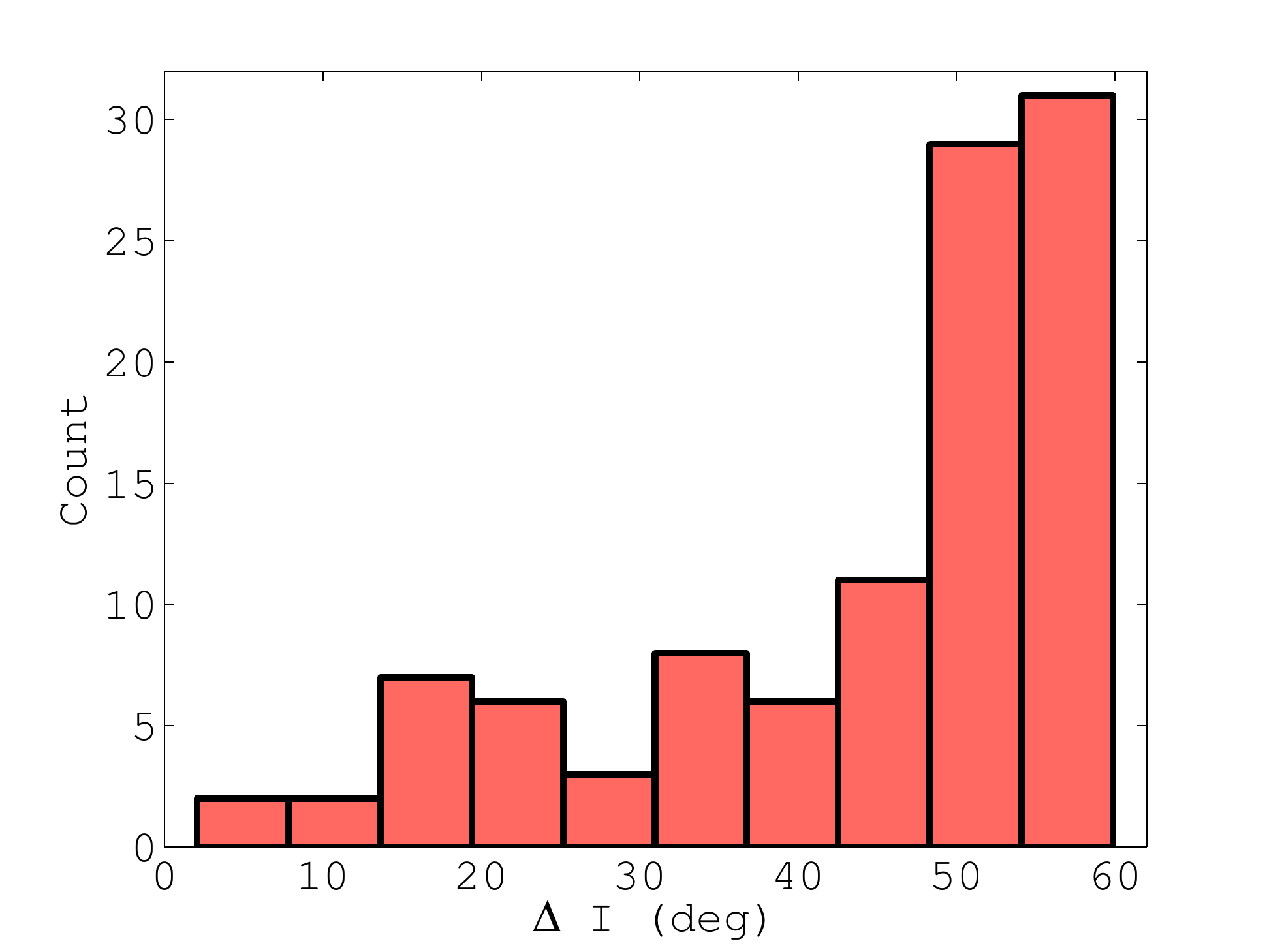}  
\caption{Histogram of the 105 systems in Cat. 3 for which transitability was predicted on either star in an eccentric system but did not occur.}

\label{fig:Error_vs_MI}
\end{center}  
\end{figure}

In Fig.~\ref{fig:Error_vs_MI} we plot a histogram of $\Delta I$ for the 105 systems in Cat. 3, for both A and B stars, where transitability was predicted but did not occur. There is a large jump at roughly $50^{\circ}$. This corresponds to the islands of libration (Fig.~\ref{fig:DoolinBlundellTest_ecc}). Such high mutual inclinations are predicted to yield transitability according to Eq.~\ref{eq:beta_factor_inequality}, but since $\Delta \Omega$ no longer passes through all angles, some unlucky aliens will never see transitability.

\begin{table*}
\caption{Testing the transitability criterion on eccentric systems} 
\centering 
\begin{tabular}{ | c | cc | cc | cc | cc | cc |} 
\hline\hline 
$\Delta I_{\rm max}$ (deg) &\multicolumn{2}{|c}{Cat. 1} &\multicolumn{2}{|c}{Cat. 2} & \multicolumn{2}{|c}{Cat. 3} & \multicolumn{2}{|c}{Cat. 4} & \multicolumn{2}{|c}{Error $\%$} \\  
\hline 
 & A & B & A & B & A & B & A & B & A & B\\
60 & 3185 & 3376 & 2677 & 2493 & 96 & 65 & 146 & 170 & 1.78 & 1.72\\
\hline 

\end{tabular}
\label{tab:EccTest}
\end{table*}

%

\end{document}